\documentclass[usenatbib]{mn2e}
\usepackage{natbibmnfix,graphicx,times, amsmath, epsfig}
\newcommand{\uunit}{{\bf \hat{u}}}
\newcommand{\cmfast}{\textsc{\small 21CMFAST}}
\newcommand{\muKK}{\mu {\rm K^2}}
\newcommand{\PkSZ}{[\Delta^{\rm patchy}_{l3000}]^2}
\newcommand{\POV}{[\Delta^{\rm OV}_{l3000}]^2}
\newcommand{\Ptot}{[\Delta_{l3000}]^2}
\newcommand{\reionparams}{\{\zeta, T_{\rm vir}, R_{\rm mfp}\}}
\newcommand{\delz}{\Delta z_{\rm re}}
\newcommand{\zre}{z_{\rm re}}

\newcommand{\avenf}{\bar{x}_{\rm HI}}

\newcommand{\lya}{Ly$\alpha$}

\newcommand{\Msun}{M_\odot}
\newcommand{\Tvir}{T_{\rm vir}}

\newcommand{\mfp}{R_{\rm mfp}}

\newcommand\lsim{\mathrel{\rlap{\lower4pt\hbox{\hskip1pt$\sim$}}
        \raise1pt\hbox{$<$}}}
\newcommand\gsim{\mathrel{\rlap{\lower4pt\hbox{\hskip1pt$\sim$}}
        \raise1pt\hbox{$>$}}}
\def\myputfigure#1#2#3#4#5%
{\vskip#5pt\makebox[0pt]{\hskip#2in
\includegraphics[width=#3\textwidth]{#1}}\vskip#4pt\hfill}

\newenvironment{packed_enum}{
\begin{enumerate}
  \setlength{\itemsep}{1pt}
  \setlength{\parskip}{0pt}
  \setlength{\parsep}{0pt}
}{\end{enumerate}}

\pdfoutput=1

\begin{document}

\title[kSZ Signal from Patchy Reionization]{
The kinetic Sunyaev-Zel'dovich signal from inhomogeneous reionization: a parameter space study\\
}

\author[Mesinger et al.]{Andrei Mesinger$^1$\thanks{email: andrei.mesinger@sns.it}, Matthew McQuinn$^2$\thanks{Einstein Fellow} \& David N. Spergel$^3$ \\
$^1$Scuola Normale Superiore, Piazza dei Cavalieri 7, 56126 Pisa, Italy\\
$^2$Berkeley Astronomy Department, University of California, Berkeley, CA 94720, USA\\
$^3$Department of Astrophysical Sciences, Princeton University, Princeton, NJ 08544, USA}

\voffset-.6in

\maketitle

\begin{abstract}
Inhomogeneous reionization acts as a source of arcminute-scale anisotropies in the cosmic microwave background (CMB), the most important of which is the kinetic Sunyaev-Zel'dovich (kSZ) effect.
 Observational efforts with the Atacama Cosmology Telescope (ACT) and the South Pole Telescope (SPT)  are poised to detect this signal for the first time, with projected 1 $\muKK$-level sensitivity to the dimensionless kSZ power spectrum around a multipole of $l=3000$, $\Ptot$.  Indeed, recent SPT measurements place a bound of $\Ptot<2.8~\muKK$ at 95\% C.L., which degrades to $\Ptot<6~\muKK$ if a significant correlation between the thermal Sunyaev-Zel'dovich (tSZ) effect and the cosmic infrared background (CIB) is allowed. To interpret these and upcoming observations, we compute the kSZ signal from a suite of $\approx 100$ reionization models using the publicly available code \cmfast.
 Our physically motivated reionization models are parameterized by the ionizing efficiency of high-redshift galaxies, the minimum virial temperature of halos capable of hosting stars, and the ionizing photon mean free path -- a parameterization motivated by previous theoretical studies of reionization.
We predict the contribution of patchy reionization to the $l=3000$ kSZ power to be $\PkSZ=$ 1.5--3.5 $\muKK$.  
  Therefore, even when adopting the lowest estimate in the literature for the post-reionization signal of $\POV \approx 2~\muKK$, none of our models are consistent with the aggressive 2$\sigma$ SPT bound that does not include correlations.
This implies that either: (i) the early stages of reionization occurred in a much more homogeneous manner than suggested by the stellar-driven scenarios we explore, such as would be the case if, e.g., very high energy X-rays or exotic particles contributed significantly; and/or (ii) that there is a significant correlation between the CIB and the tSZ.  The later is perhaps not surprising, as massive halos should host both hot gas and star forming galaxies.
On the other hand, the conservative SPT bound of $\Ptot \lsim 6~\muKK$ is compatible with all of our models and is on the threshold of constraining physically motivated reionization models.  The largest patchy kSZ signals correspond to an extended reionization process, in which the sources of ionizing photons are abundant and there are many recombinations (absorptions in sinks).
We point out that insights into the astrophysics of the early Universe are encoded in both the amplitude and shape of the kSZ power spectrum.
\end{abstract}


\section{Introduction}
\label{sec:intro}

Scattering of cosmic microwave background (CMB) photons off of inhomogeneities in the electron density
after the recombination epoch imprints ``secondary'' anisotropies in the CMB.  These secondaries are the dominate contribution to the CMB anisotropies at $\lsim 5'$ scales.   The largest and most studied
of these secondaries is the thermal Sunyaev-Zel'dovich (tSZ) effect, which results from Compton scattering off of $\sim 10^7~$K
intracluster gas \citep{Zeldovich69}, and most of the tSZ arises from $z<1$ (e.g., \citealt{Komatsu02}).  The second most prominent is the kinetic Sunyaev-Zel'dovich (kSZ) effect and is the subject of this study.  In contrast to the tSZ, the kSZ is not sourced by rare regions in the Universe but arises because most of the volume has a bulk peculiar motion with respect to the CMB frame \citep{SZ80, OV86, Vishniac87, MF02}.  These motions impart a Doppler shift to the $\approx 10\%$ of CMB photons that scatter after recombination.  The kSZ is typically generated at higher redshifts than the tSZ, with a large fraction likely being sourced by inhomogeneities at $z\gsim 6$, during the epoch of reionization \citep{GH98, KSD98, VBS01, Santos03, Salvaterra05, McQuinn05, Zahn05}.

Until recently, the prospects for detecting the kSZ were perceived by the CMB community as bleak.  While the unique frequency dependence of the tSZ anisotropies should allow the separation of the kSZ from the tSZ, theoretical projections were that the tSZ signal would be more than an order of magnitude larger than that of kSZ on arcminute scales (e.g., \citealt{Zahn05}).  As such, instruments targeting the CMB secondaries would face the difficult task of using a limited number of frequency channels ($\sim 3$) to isolate the much smaller kSZ signal while also using this frequency information to isolate the contamination from dusty galaxies and active galactic nuclei \citep{HS05}.  However, recent measurements have indicated that the amplitude of the tSZ power spectrum may actually be comparable to that of the kSZ rather than an order of magnitude larger.  This lower value has been attributed to the lower $\sigma_8$ that is preferred by current cosmological measurements relative to the estimates of a few years ago (the amplitude of the tSZ power spectrum scales as $\sim \sigma_8^9$; \citealt{Komatsu02}), and also to the realization that feedback processes strongly suppress the tSZ in clusters \citep{Battaglia10, Shaw10, TBO11}.  As a result, separating the two signals is now expected to be an easier task than had been anticipated.

Both the Atacama Cosmology Telescope (ATC)\footnote{http://www.physics.princeton.edu/act/} and South Pole Telescope (SPT)\footnote{http://pole.uchicago.edu/} began measuring the CMB at arc-minute scales in 2007 and have recently published upper limits on the kSZ angular power spectrum (defined in Section 2).
At present, the greatest challenge for these efforts is separating out several different terms that contribute to the angular power spectrum: the primordial CMB, the tSZ, radio sources, the cosmic infrared background (CIB), in addition to cross-power between the tSZ and radio sources, and between the tSZ and the CIB.   The primordial fluctuations are important at low multiples and the sources (CIB/dusty galaxies and radio sources) are most important at high multipoles. The tSZ-CIB cross-correlation is currently the least understood component.

 With ACT's  two frequency bands, \citet{Dunkley11} placed a limit of 
${\it l}^2/[2\pi] [ C^{\rm tSZ}_l + C^{\rm kSZ}_l] =  6.8\pm 2.9~\muKK$ at 
 the angular multipole of $l \approx 3000$,  where $C^{\rm tSZ}_l$ and  $C^{\rm kSZ}_l$ are the tSZ and kSZ angular power spectra, respectively.  Using  SPT's three frequency bands and ignoring the tSZ-dusty galaxy correlation,  \citet{Reichardt11} recently reported an upper bound ${\it l}^2/[2\pi] \,C^{\rm kSZ}_l < 2.8~\muKK$ at 95\% C.L..
 \citet{Reichardt11} did note that the kSZ constraint is significantly weakened to ${\it l}^2/[2\pi] \,C^{\rm kSZ}_l< 6~\muKK$ if this study allows for a significant anti-correlation between the tSZ and the cosmic infrared background.   
Analysis of the Planck data \citep{Planck18} suggest that 90\% of the dust anisotropy power at $l \sim 3000$ is from $z > 2$ while simulations of the tSZ signal predict that the bulk of the signal is sourced from lower redshifts \citep{Battaglia11,TBO11}.  While this Planck analysis argues that the tSZ-CIB cross-correlation is small, further work is needed to better characterize the redshift distribution of these signals as current models do not provide good fits to the Planck data \citep{Shang11}.  The detection of the kSZ signal will likely require a combination of Herschel measurements of the dusty galaxies and multi-frequency millimeter observations by SPT and ACT. Nevertheless, even the recent conservative bound of ${\it l}^2/[2\pi] \,C^{\rm kSZ}_l< 6~\muKK$  rules out the most extended reionization models in \citet{McQuinn05}.

The kSZ is comprised of two components: a component sourced primarily after reionization and referred to as the Ostricker-Vishniac (OV) effect \citep{OV86, MF02}\footnote{In some prior studies, ``OV'' has been used to refer to the linear-theory, post-reionization kSZ and the nonlinear theory was referred to as the ``nonlinear kSZ.''  We do not adopt this terminology here.} and a component from during reionization referred to as the patchy kSZ \citep{GH98, KSD98}.   The OV effect results from scattering off of density inhomogeneities with bulk peculiar velocities, with the fractional contribution per unit redshift peaking at $z\sim 1$ and monotonically decreasing with redshift above this peak.  The OV is sourced by relatively linear scales, and it is likely that it can be modeled well enough to isolate the kSZ contribution from reionization.   The patchy kSZ signal results from the order unity fluctuations in the ionized fraction that are anticipated by the vast majority of reionization models.  
   Reionization also imprints other anisotropies in the CMB, primarily in polarization \citep{Hu00, Dore07, DS09, Su11}, which are not discussed here. However, other than the large-scale ($l \lsim 10$) polarization anisotropies, the kSZ is by far the most detectable of the reionization-induced anisotropies.

The patchy kSZ anisotropies were originally calculated for
several toy models of patchy reionization \citep{GH98, KSD98, VBS01, Santos03}.  In the last several years, the kSZ from reionization has been calculated for more physically motivated models of reionization either using numerical simulations or semi-analytic models \citep{Zahn05, McQuinn05, Iliev07kSZ}.   Numerical simulations of reionization have not achieved large enough box sizes to both resolve this process and also capture the velocity flows which are correlated over hundreds of Mpc and which contribute significantly to the amplitude of the kSZ anisotropies (although, see \citealt{Iliev07kSZ} for a method to account for flows larger than the box size).  More fundamentally, the heavy computational requirements of numerical simulations do not allow for exploration of the large astrophysical parameter space associated with reionization.  Because of these limitations, recent years have seen the development of so called ``semi-numerical'' simulations of reionization, which incorporate an excursion-set model of reionization \citep{FHZ04} with large-scale realizations of the density field to produce 3D reionization morphology \citep{Zahn05, MF07, GW08, Alvarez09, CHR09, MFC11, Zahn11}.
In this study, we use the publicly available \cmfast\ code\footnote{http://homepage.sns.it/mesinger/Sim.html} to generate $\sim$ 100 reionization scenarios in order to explore the parameter space of the patchy kSZ signal \footnote{When this work was nearing completion, a complementary study by \citet{Zahn12} appeared in the literature.  This study focused on constraining the duration and timing of reionization with recent SPT results.  We will briefly compare our results to theirs in \S \ref{sec:prior}.}.

There are many other potential probes of reionization, including: the CMB Thompson scattering optical depth, $\tau_e$, (e.g., \citealt{Komatsu11}); the redshifted 21cm line from neutral hydrogen (see, e.g., \citealt{FOB06} and references therein); transmission statistics of quasar (QSO) spectra (e.g., \citealt{WL04_nf, MH04, MH07, Fan06, BH07_quasars, Maselli07, Gallerani08, MMF11, Bolton11});
 intergalactic damping wing absorption in gamma-ray burst (GRB) spectra \citep{Miralda-Escude98, BL04_grbvsqso, Totani06, McQuinn08, MF08damp}; the number density, clustering and evolution of Ly$\alpha$ emitters (LAEs; \citealt{MR04,Santos04, HC05,FZH06,MR06, Kashikawa06,McQuinn07LAE, DWH07,MF08LAE,Iliev08, DMW11}). 
However the CMB-based constraints  
are likely to be the only ones in the next decade that are capable of probing the highest-redshift part of reionization, $z\gsim10$.

This paper is organized as follows. In \S \ref{sec:kSZ}, we summarize our method for estimating the kSZ signal.  Our main results are presented in \S \ref{sec:results}, namely the kSZ signal from a large range of reionization scenarios. In \S \ref{sec:prior} we compare our results to previous works.  In \S \ref{sec:detect}, we put our results in context with recent and forthcoming observations of the kSZ. Finally in \S \ref{sec:conc} we present our conclusions.

Unless stated otherwise, we quote all quantities in comoving units. We adopt the background cosmological parameters \{$\Omega_\Lambda$, $\Omega_{\rm M}$, $\Omega_b$, $n$, $\sigma_8$, $H_0$\} = \{0.73, 0.27, 0.046, 0.96, 0.82, 70 km s$^{-1}$ Mpc$^{-1}$\}, consistent with the seven--year results of the {\it WMAP} satellite \citep{Komatsu11}.   We use the Fourier convention that is the standard in modern cosmological studies in which the $2\pi$'s appear under the $dk$'s.

\section{Estimating the kSZ signal}
\label{sec:kSZ}

The CMB temperature anisotropy in the direction of the line-of-sight unit vector, $\uunit$, can be written as: 
\begin{equation}
\label{eq:kSZ}
\delta_T \equiv \frac{\Delta T}{T}(\uunit) = \sigma_{\rm T} \int dz \,c\, (dt/dz) \, e^{-\tau_e(z)} n_e {\bf \uunit \cdot v}  ~ ,
\end{equation}
\noindent where $\sigma_{\rm T}$ is the Thomson scattering cross section, $\tau_e(z)$ is the Thomson optical depth to redshift $z$ in the direction $\uunit$, ${\bf v}(\uunit, z)$ is the peculiar velocity, and $n_e(\uunit, z)$ is the electron number density.

In our investigations, we split up the integral in eq. (\ref{eq:kSZ}) into the post-reionization, low-redshift component -- the OV signal -- and the high-redshift, patchy kSZ reionization component.  The patchy kSZ signal is defined in this study as the total kSZ from $z>5.6$, even in cases for which reionization completed at redshifts higher than $z=5.6$.  However, the patchy reionization signal typically dominates the total kSZ from $z>5.6$ and so this somewhat arbitrary definition makes little difference and facilitates the comparison of models.  While the patchy kSZ signal is the focus of this paper, we discuss both the patchy and OV components in turn below.

We focus on characterizing the kSZ angular power spectrum $C_{l}^{\rm kSZ} \equiv T_{\rm cmb}^2 |\tilde{\delta_T}(k)|^2$, where $\tilde{\delta_T}$ is the Fourier transform of $\delta_T$ and $T_{\rm cmb} = 2.73~$K is the temperature of the CMB.  This statistic is the final analysis product of the relevant CMB observations.  We will use the notation $[\Delta_{l}^{\rm kSZ}]^2 \equiv l^2 C_{l}^{\rm kSZ}/(2\pi)$.  
  It is intuitive to relate in the flat sky approximation $C_{l}^{\rm kSZ}$ to the $3$D ionization and density power spectra:
\begin{eqnarray}
C_{l}^{\rm kSZ} &\approx & \frac{T_{\rm cmb}^2}{3}\int \frac{d \eta}{\eta^2} \, a^{2} \, c^{-2} v_{\rm rms}(z)^2 \, \sigma_{\rm T}^2 \, \bar{n}_{\rm e}(z)^2 \, \bigg[P_{xx}(z, \frac{l}{\eta}) \nonumber \\
&+&  2 \,P_{x \rho}(z, \frac{\bf l}{\eta}) 
 +   P_{\rho \rho}(z, \frac{ l}{\eta}) \bigg],
 \label{eqn:ctau}
\end{eqnarray}
where we have also approximated the cosmological velocity flows as coherent over much larger scales than those of interest, $\eta/l$ where ${\it l} \approx 3000$ (e.g., \citealt{MF02}).  Here, $\eta$ is the conformal distance from the observer, $v_{\rm rms}(z)$ is the RMS amplitude of the peculiar velocity, 
and $P_{XY}$ is the 3D cross power spectrum of the overdensity in $X$ with the overdensity in $Y$, where the relevant overdensities are in the ionized fraction, $x$, and the density, $\rho$.  During reionization the patchy term dominates, i.e., the term that arises from $P_{xx}$.  In the absence of redshift evolution in the power spectra and the electron fraction, the integral in equation (\ref{eqn:ctau}) grows as $(1+z_{\rm max})^{3/2}$ during matter domination, where $z_{\rm max}$ is the maximum integrated redshift.  Thus, the same ionization topology at $z=10$ contributes slightly more to the kSZ anisotropy than this topology at $z=7$.  More importantly, doubling the duration of the epoch when ionization structures are of a given scale 
 approximately doubles the patchy kSZ signal, $C_{l}^{\rm patchy}$, at that scale.

As a final remark, we have argued that reionization almost certainly occurred in a patchy manner, consisting of ionized and neutral regions.  Even without any knowledge of $P_{xx}$ other than that reionization was patchy -- composed of zeros and ones --, there is an integral constraint on the reionization contribution to $C_{l}^{\rm kSZ}$ because the variance of the ionized fraction (or $x_i^{-1} -1$, where $x_i$ is the ionized fraction) must equal $\int k^2dk/(2\pi^2)  \, P_{xx}$.  Thus, fixing the reionization history, $ \int dl \, [\Delta_{l}^{\rm patchy}]^2$ is a single number that is independent of the reionization morphology.  This constraint suggests that smaller bubbles result in a smaller signal since $[\Delta_{l}^{\rm patchy}]^2 \sim~({\rm characteristic~\it l})^{-1}\sim$~bubble~size.  However, the clustering of bubbles can invalidate this simple interpretation.

\subsection{kSZ from patchy reionization}
\label{sec:reion_sim}

\begin{figure*}
\vspace{+0\baselineskip}
{
\includegraphics[width=\textwidth]{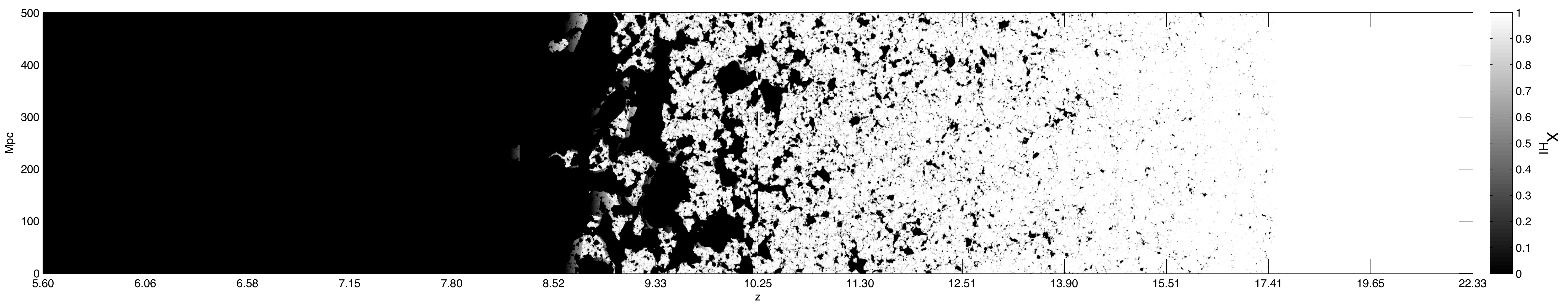}
\includegraphics[width=\textwidth]{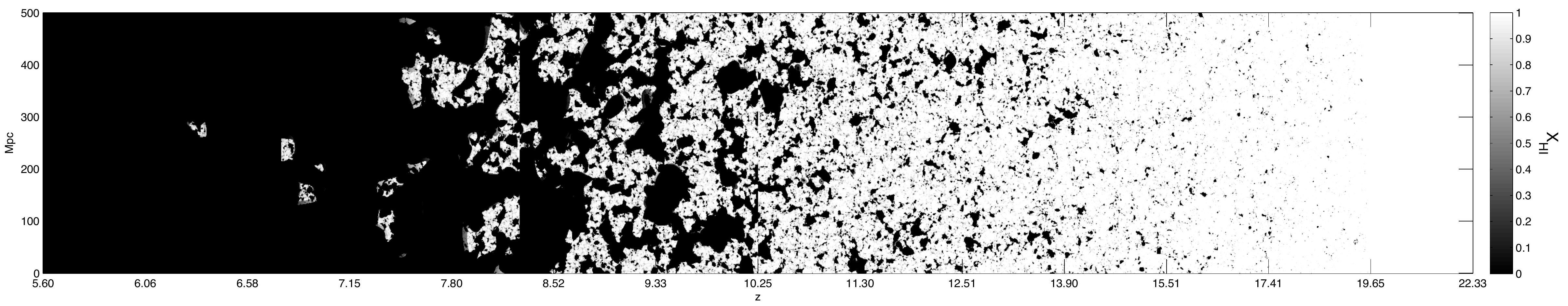}
\includegraphics[width=\textwidth]{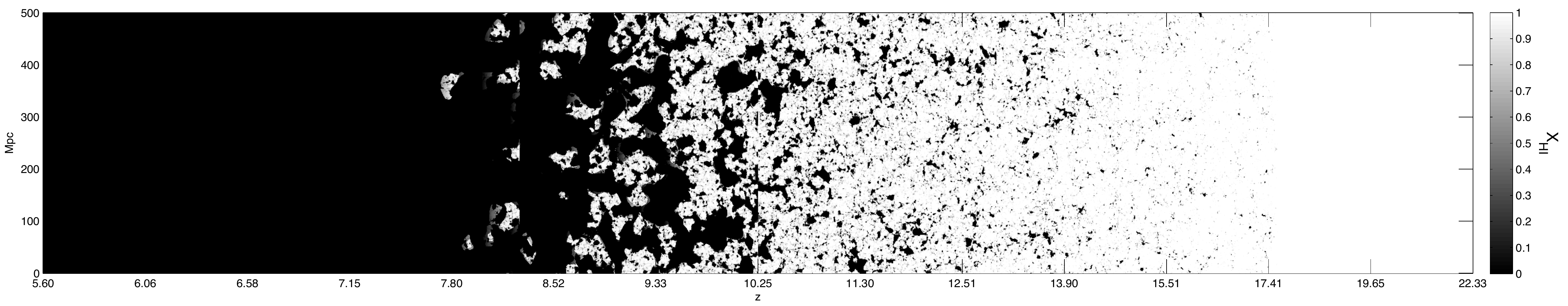}
\includegraphics[width=\textwidth]{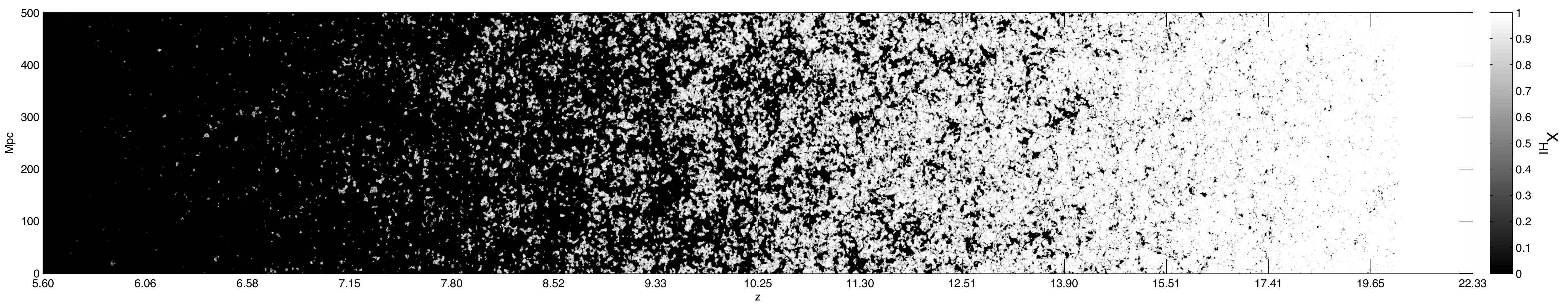}
}
\caption{
Slices through our ionization fields with thickness 1.1 Mpc.
  The top panel corresponds to $\reionparams$ = \{32, $10^4$ K, 60 Mpc\}.  The middle two panels illustrate the impact of temporal evolution in our astrophysical parameters: $\reionparams$ = \{$32 [(1+z)/11]^2$, $10^4$ K, 60 Mpc\}, and $\reionparams$ = \{32,  $10^4$ K, $60 \, [7/(1+z)]^3$ Mpc\}, from top to bottom.  The bottom panel corresponds to an extreme scenario with a small mean free path, $\reionparams$ = \{32, $6.3\times10^3$ K, 3 Mpc\}.  The models have $\tau_e$= 0.087, 0.086, 0.085, and 0.098, consistent at $1\sigma$ with {\it WMAP}, and $\PkSZ$= 2.4, 3.3, 2.5, and 3.1 $\muKK$ ({\it top to bottom, respectively}).  The vertical lines that are sometimes evident (such as in the second panel at $z\approx8.3$) owe to how the snapshots are stacked (see text for details).
\label{fig:ionization_maps}
}
\vspace{-1\baselineskip}
\end{figure*}


To model the patchy kSZ signal, we use the publicly available \cmfast\ code\footnote{http://homepage.sns.it/mesinger/Sim.html}. This code uses perturbation theory (PT) to generate the density and velocity fields, and PT plus the formalism of \citet{FZH04} to generate the ionization fields.  
We summarize these calculations below.  For further details on these algorithms, see \citet{MF07} and \citet{MFC11}.

We first create a Gaussian random field in a simulation box which is 500 Mpc on a side, sampled onto a 1800$^3$ grid.  This field is then mapped onto a lower-resolution 450$^3$ Eulerian grid at a given redshift using first-order Lagrangian PT \citep{Zeldovich70} to compute what we refer to as the ``nonlinear density field'' and corresponding velocity field.  The statistical properties of the density and velocity fields have been shown to match those from a hydrodynamic simulation remarkably well over the range of scales and redshifts relevant for patchy kSZ measurements \citep{MFC11}. 

Our ionization fields are generated directly from the non-linear density field, with the excursion-set prescription of \citet{FZH04}, slightly modified according to \citet{Zahn11} (see their 'FFRT' scheme).  This algorithm marks regions as ionized for which the number of ionizing photons produced is greater than the number of neutral atoms.
Specifically, a simulation cell at coordinate ${\bf x}$ is flagged as ionized if 
\begin{equation}
\label{eq:HII_barrier}
f_{\rm coll}({\bf x}, z, R) \geq \zeta^{-1} ~ ,
\end{equation}
\noindent where $\zeta$ is an ionizing efficiency parameter (described later) and $f_{\rm coll}$ is the fraction of mass residing in dark matter halos inside a sphere of radius $R$ and mass $M=4/3 \pi R^3 \rho$, where $\rho = \bar{\rho} [1+\langle \delta_{\rm NL} \rangle_R]$.  This tabulation includes all halos above a set virial temperature threshold, $\Tvir$.  We do not explicitly compute individual halo locations (as in \citealt{MF07}), but instead use the non-linear density field when computing $f_{\rm coll}$ \citep{Zahn11}.   Following the excursion-set approach (e.g., \citealt{Bond91, LC93}), we iteratively decrease the scale $R$, starting from some maximum value, $R_{\rm max}$, which we specify below as $R_{\rm mfp}$.  
 If at any $R$ the criterion in equation (\ref{eq:HII_barrier}) is met, this cell is flagged as ionized.
We also take into account partially ionized cells by setting the cell's ionized fraction to $\zeta f_{\rm coll}({\bf x}, z, R_{\rm cell})$ at the last filter step for those cells which are not fully ionized.  The resulting ionization fields are found to agree well with those generated with cosmological radiative transfer algorithms on scales relevant for the kSZ \citep{Zahn11, MFC11}.

We parameterize reionization with $3$ free parameters:
\begin{packed_enum}
\item $\zeta$: {\it the ionizing efficiency of high-redshift galaxies.}  This quantity can be defined as $\zeta = f_{\rm esc} f_\ast N_\gamma / (1+n_{\rm rec})$, where $f_{\rm esc}$ is the fraction of ionizing photons produced by stars that escape into the intergalactic medium (IGM), $f_\ast$ is the star formation efficiency, $N_\gamma$ is the number of ionizing photons per stellar baryon, and $n_{\rm rec}$ is the mean number of recombinations per baryon.  For reference, $f_{\rm esc}=0.1$, $f_\ast=0.1$, $N_\gamma =4000$ (appropriate for PopII stars), and $n_{\rm rec}=1$ yield $\zeta = 20$.  However, the parameters $f_{\rm esc}$ and  $f_\ast$ are extremely uncertain in high-redshift galaxies  (e.g., \citealt{GKC08, WC09, Paardekooper11}). In this work, we explore the range $\zeta\sim$ 10-50, which corresponds to what is consistent with CMB and Ly$\alpha$ forest constraints on reionization.\footnote{Simple models at moderate redshifts find that a linear scaling between halo mass and UV luminosity, like the one used here (albeit for lower halo masses), provides a good fit to the observed galaxy luminosity function and their clustering properties (e.g., \citealt{VO96, Lee09}).  Nevertheless, it is possible that the ionizing efficiency of galaxies varies significantly with halo mass at high redshifts.  Since the dominant ionizing population at early times likely corresponds to a narrow range in halo masses (because of the steepness of the halo mass function), a halo mass-independent value of $\zeta$ may still be a good approximation.}

\item $\Tvir$: {\it the minimum virial temperature of halos that host the reionization sources.} Much of the reionization literature sets this to $\Tvir\approx10^4$ K, the threshold temperature for efficient atomic cooling  -- corresponding to a halo mass of $M_{\rm halo}\sim 10^8\Msun$ at $z\sim10$.  However, $\Tvir$ could be smaller than $10^4$K:
The first stars were likely hosted by smaller halos with $M_{\rm halo}=10^{6-7}\Msun$ (e.g., \citealt{HTL96, ABN02, BCL02}).  However, star formation inside such small halos was likely inefficient (with a handful of stars per halo), and was eventually suppressed by the heating from reionization itself or other feedback processes \citep{HAR00,RGS01,MBH06, HB06}. 
  $\Tvir$ could also have been larger than $10^4$K:  Radiative and/or mechanical feedback (e.g., \citealt{SH03}) eventually suppresses star formation inside $\Tvir \lsim 10^5$ K halos, though the details and timing of these processes are not well understood at high redshifts (e.g., \citealt{MD08, OGT08, PS09})\footnote{Note that the combination of mechanical and radiative feedback, which by themselves likely have different mass-scalings, could conspire to create a relatively flat (i.e., with weak halo mass dependence) suppression of star formation (e.g., \citealt{FDO11}).  Such an effect most appropriately translates to a lower value of $\zeta$ (e.g., by lowering $f_\ast$).}.
Nevertheless, it is unlikely that $\Tvir$ is much greater than few$\times10^5$~K, since these values approximately latch onto the faint end of the observed galaxy luminosity functions at $z\sim6$ (e.g., \citealt{Bouwens08, Labbe10, SFD11, FDO11}).  Additionally, high $\Tvir$ models have difficulty in latching onto the slow evolution of the observed emissivity, as inferred from the \lya\ forest at $3<z<6$ (see Fig. \ref{fig:emissivity} and associated discussion). Thus, we explore the range $\Tvir=10^3$--$3\times10^5$ K, which encompasses all conceivable $\Tvir$.


\item $\mfp$: {\it the ionizing photon mean free path within ionized regions of the IGM.}  The mean free path to intersect an optically thick system (termed Lyman Limit Systems; LLSs) is measured to have been $\mfp\sim $ 50 comoving Mpc at $z= 6$, albeit with order-unity uncertainty (e.g., \citealt{Storrie-Lombardi94, Stengler-Larrea95, Peroux03, PWO09, SC10}).
   This parameter becomes important in our models when the ionized region size $R$ is larger than $\mfp$, resulting in most of the ionizing photons being absorbed in LLSs rather than in the diffuse IGM, thereby retarding reionization.  Thus, the smallest values for $\mfp$ have the largest impact in our models.\footnote{Note that in our simple model, the mean free path is spatially homogeneous.  There are likely inhomogeneities in $\mfp$ during reionization, though the extent of these inhomogeneities is currently unknown (e.g., \citealt{CHR09, Crociani11, MOF11}).}
 In this work, we explore the range $\mfp=$ 3--80 Mpc.  
\end{packed_enum}

Our parameterization ignores redshift evolution of the above astrophysical parameters.  Since reionization is expected to have been extended, it is in fact likely that there would have been some evolution in these parameters.  However, we find that only a faction of the reionization history contributes the bulk of the patchy kSZ signal, making evolution less important.
 Nevertheless, we also investigate a few scenarios that have evolving parameters. 

In order to explore the patchy kSZ signal in this $3$ parameter volume, we have simulated 103 different reionization models.  The ionization boxes for each model are computed at redshift intervals of $\Delta z=0.2$, spanning the redshift range $z=$ 5.6--22.4. This procedure generates 85 ionization boxes for each parameter combination, with the entire $\approx$100-point parameter space being sampled in roughly one week on an 8-core desktop computer.
We interpolate the density, velocity and ionization fields linearly in cosmic time between two consecutive simulation outputs.
  As is traditionally done in secondary CMB anisotropy calculations, we rotate the boxes when reaching their edges so as to minimize coherent stacking of the signal.  Sample slices through four simulations are shown in Fig. \ref{fig:ionization_maps}.
  We have performed box size and resolution tests with a fiducial model and found convergence in the $l\sim$2000-10000 kSZ power to $\approx$ 10\% (see Appendix A).

\subsection{The post reionization kSZ} 
\label{sec:OV}

\begin{figure}
\vspace{-1\baselineskip}
\includegraphics[width=0.45\textwidth]{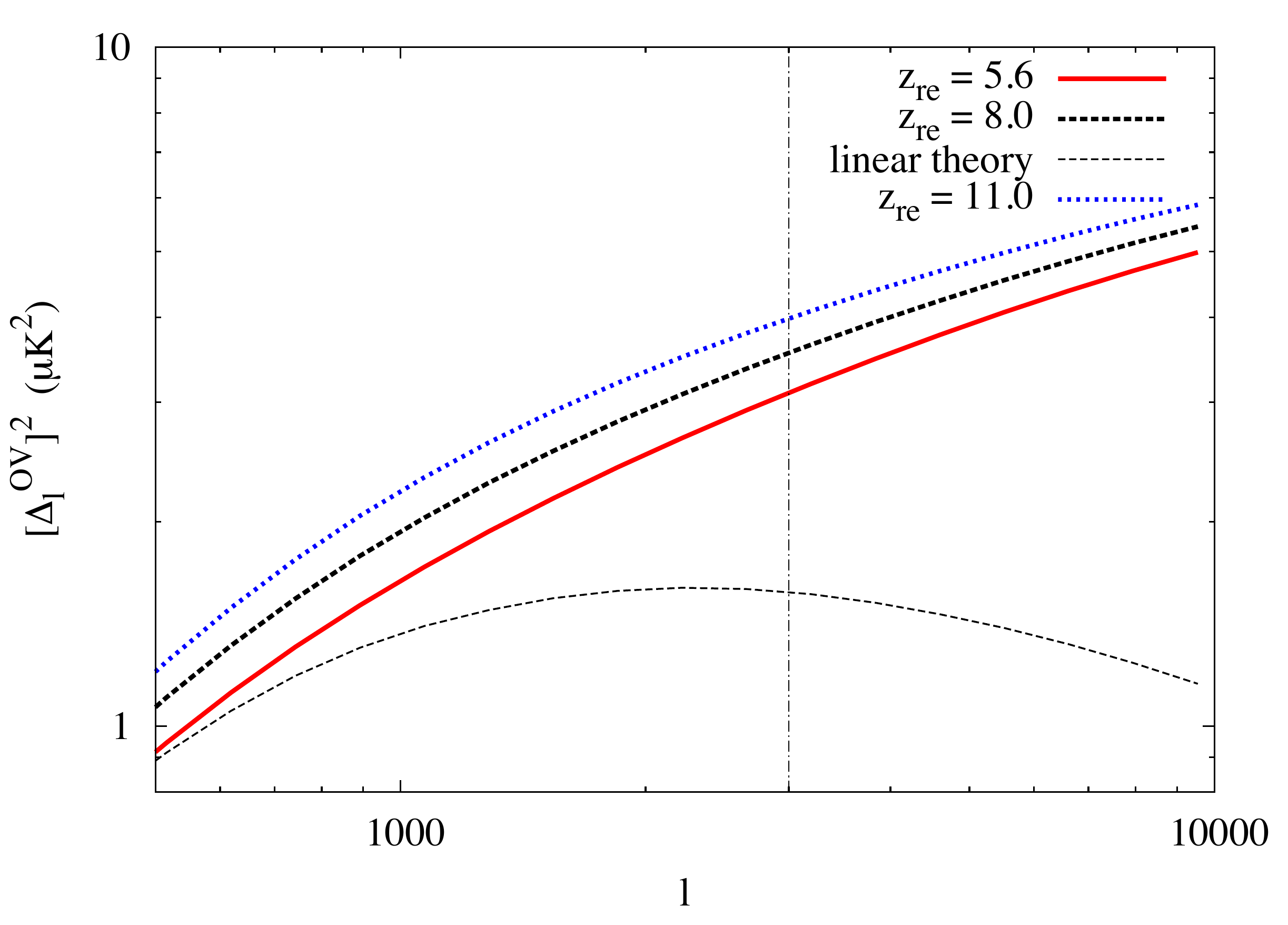}
\caption{
OV dimensionless angular power spectrum for instantaneous reionization scenarios with a reionization redshift of $z_{\rm re} = 5.6$ (solid curve), $8$ (dashed), and $11$ (dotted).  We have also plotted the OV power spectrum calculated in linear theory for the $z_{\rm re} =8$ case.
\label{fig:OV}
}
\vspace{-1\baselineskip}
\end{figure}

The kSZ from after reionization -- which we refer to 
 as the OV effect -- is a contaminate for studying the patchy kSZ signal.  Any modeling uncertainty in this post-reionization component will bias constraints on the patchy component. At $l\approx3000$, the OV anisotropy contributed per unit redshift peaks at $z\approx 1$ and falls off weakly towards higher redshifts (becoming zero when there are no longer free electrons around).  Figure \ref{fig:OV} shows the OV anisotropies for three instantaneous reionization scenarios with reionization redshifts of $z_{\rm re} = 5.6$, $8$, and $11$.  We find the amplitude of the OV dimensionless (i.e. without units of length) angular power spectrum at $l = 3000$ to be $\POV =$ 3--4 $\muKK$.  
 
 These curves were calculated using the pseudo-nonlinear formula for the OV derived in \citet{MF02} (a more accurate expression than in eq. \ref{eqn:ctau}).  This calculation takes the linear theory OV calculation of \citet{Vishniac87} and substitutes a nonlinear theory for the density field, here computed from the Peacock and Dodds density power spectrum \citep{PD94}.\footnote{In our calculation, we use linear perturbation theory for the velocity field, an approximation justified in \citealt{MF02}. In addition, we note that the density power spectrum has not been ``filtered'' in our calculations to account for the pressure smoothing of the baryons (as was done in \citealt{McQuinn05} and references therein), but note that this should have little effect at the multipole of interest, $l = 3000$, unless there is significant feedback.  We verified this for the window function measured in \citealt{SRN11}, e.g. their eq. (21), which we find suppresses the power at $l = 3000$ by only a few percent.}  At higher multipoles than $l = 3000$, this suppression is more significant.  For reference, $\eta  \, 2\pi/{l}  = \{4, ~7,~ 14,~ 20\}~$Mpc at $z= \{0.5,~ 1,~3,~ 10\}$ at $l = 3000$.  The fact that these are pseudo-linear scales is why the linear theory calculation only differs from the nonlinear calculation by a factor of $\approx 2$ (compare the dashed curves in Fig. \ref{fig:OV}, which correspond to the linear and nonlinear OV for $z_{\rm re} = 8$).
This relatively small difference suggests that the modeling uncertainty is less severe for the OV compared to, for example, the highly nonlinear thermal SZ. 

How do our analytic predictions for the OV compare with those in numerical simulations?  As mentioned in the introduction, simulating the kSZ is notoriously difficult (e.g., \citealt{ZPT04}), and most numerical calculations of the OV should be considered lower limits on this signal owing to missing large-scale velocity modes.  However, some recent simulations have achieved box sizes where they should be capturing the large-scale velocity flows.  Of note, \citet{TBO11} investigated the OV assuming a similar cosmology as is used here and found a signal that ranged between $\POV = 1.8-2.4\,\muKK$ at $l = 3000$ for $z_{\rm re} = 10$ in the four scenarios they investigated. (The lowest value in this range was from a simulation with a strong feedback prescription.)  A similar range of values was found in \citet{SRN11}, who argued that baryonic physics/feedback could suppress the kSZ by a maximum of 1~$\muKK$ from our fiducial estimates that do not include such effects.  We adopt this number as the maximal suppression in the OV from our calculations, i.e., $\POV >$ 2 $\muKK$. 

{\it We henceforth only plot the kSZ power that originates from $z >5.6$, which is dominated by the patchy component of the kSZ.}  We have argued that the $z <5.6$ contribution has $\POV\equiv 3000^2 \,C^{\rm OV}_{l=3000} /2/\pi \approx$ 2--3 ~$\muKK$  at $l=3000$.  This range is one-third to half of the conservative $<6~{\rm \mu K}^2$ bound of SPT \citep{Reichardt11}, resulting in the constraint $\PkSZ \equiv 3000^2 \, C^{\rm patchy}_{l=3000} /2/\pi < 3-4 \muKK$.  However, the more aggressive bound of $\Ptot < 3~{\rm \mu K}^2$ from \citealt{Reichardt11} would imply $\PkSZ < 1~ \muKK$.

\section{Results}
\label{sec:results}

\subsection{Power spectra and physical intuition}
\label{sec:ps}

In this section we present the kSZ signal is several of our models.  Qualitatively a few trends are anticipated.  First, early and extended reionization scenarios should have a larger kSZ signal.  Second, the peak in ionization power shifts from small to large scales as reionization progresses and the HII bubbles grow (e.g., \citealt{McQuinn07}).  Therefore the kSZ power at higher multipoles probes earlier epochs (see Fig. \ref{fig:ps_breakdown}).

\begin{figure}
\vspace{-1\baselineskip}
\includegraphics[width=0.45\textwidth]{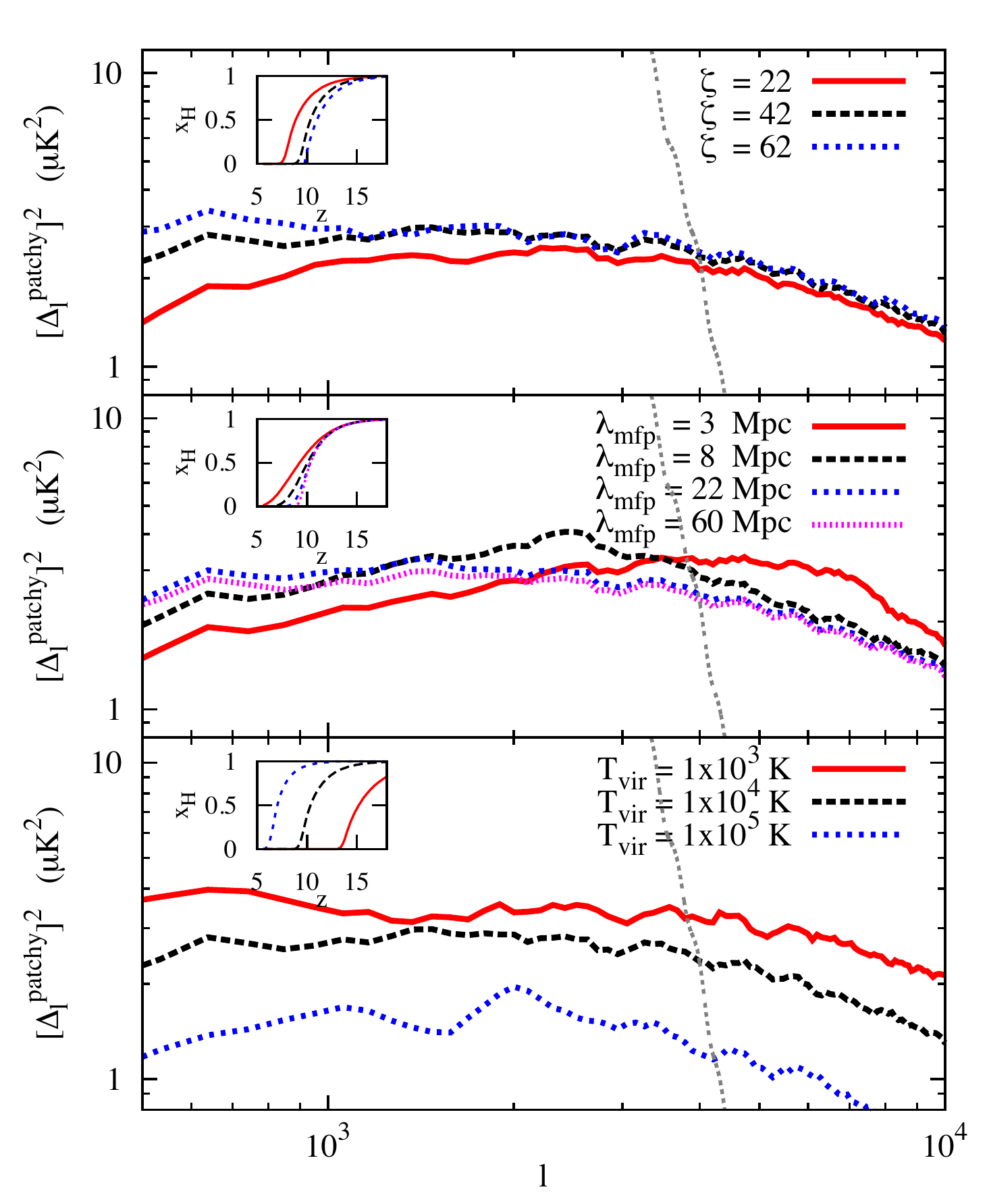}
\caption{
The dimensionless angular power spectrum of the patchy kSZ, $[\Delta^{\rm patchy}_{l}]^2$, as $\zeta$, $\mfp$, and $\Tvir$ are separately varied around $\reionparams$ = \{42, $10^4$ K, 60 Mpc\}, (top, middle, and bottom panels, respectively). The inset in each panel shows the associated reionization history.  The near-vertical gray dashed line corresponds to the primary CMB anisotropies (which are understood well enough that their contribution can be isolated at ${\it l} \gsim 2500$). 
\label{fig:ps_params}
}
\vspace{-1\baselineskip}
\end{figure}

To quantify these trends, we begin by showing the patchy kSZ angular power spectra for a sample of reionization models.  Fig. \ref{fig:ps_params} shows how the kSZ power spectra change as we vary $\zeta$ ({\it top panel}), $\mfp$ ({\it middle panel}), and $\Tvir$ ({\it bottom panel}).  The inset in each panel shows the associated reionization history.  Unless specified otherwise, the parameter values for each curve are $\reionparams$ = \{42, $10^4$ K, 60 Mpc\}.

Increasing the ionizing efficiency parameter, $\zeta$, quickens the progress of reionization as galaxies are more efficient ionizers of the IGM.  The differences in the reionization histories ($\avenf$ at fixed $z$) increase with time (decreasing redshift).  Therefore changing $\zeta$ has the most significant impact on the smallest multipoles (larger spatial scales) that are shown, since these multiples are sourced by the later stages when ionized structures were large.

Changing the mean free path of ionizing photons, $\mfp$, primarily impacts the middle and late stages of reionization.  Small values of $\mfp$ extend the end of reionization as seen in the inset in the middle panel of Fig. \ref{fig:ps_params} (c.f. \citealt{FM09, AA10}).  When the size of the ionized bubbles approaches $\mfp$, reionization is slowed as ionizing photons become increasingly swallowed by photon sinks instead of pushing out the edges of the HII regions.  Remaining neutral regions must then be ionized by local sources.
  We see that for this fiducial model where atomically cooled
 halos host ionizing sources, the mean free path only affects the kSZ signal when $\mfp \lsim 15$ Mpc.  As $\mfp$ is decreased (particularly to the smallest values considered), the peak of the patchy kSZ power spectra shifts to smaller scales.\footnote{One might worry that $\mfp=3$ Mpc is only a factor of $\approx3$ larger than our cell size.  To test whether this resolution limit has a quantitative impact on our results, we ran a higher resolution, $750^3$, run with the same astrophysical parameters as the model featured in the bottom panel of Fig. \ref{fig:ionization_maps}, and found good convergence (see Appendix A).}

Increasing $\Tvir$ delays reionization until the halos above this virial temperature threshold can form in sufficient abundance (bottom panel, Fig. \ref{fig:ps_params}).  Increasing $\Tvir$ also results in a more sudden reionization as the number of halos with time grows more quickly as the halos become rarer.

\begin{figure}
\vspace{-1\baselineskip}
\includegraphics[width=0.45\textwidth]{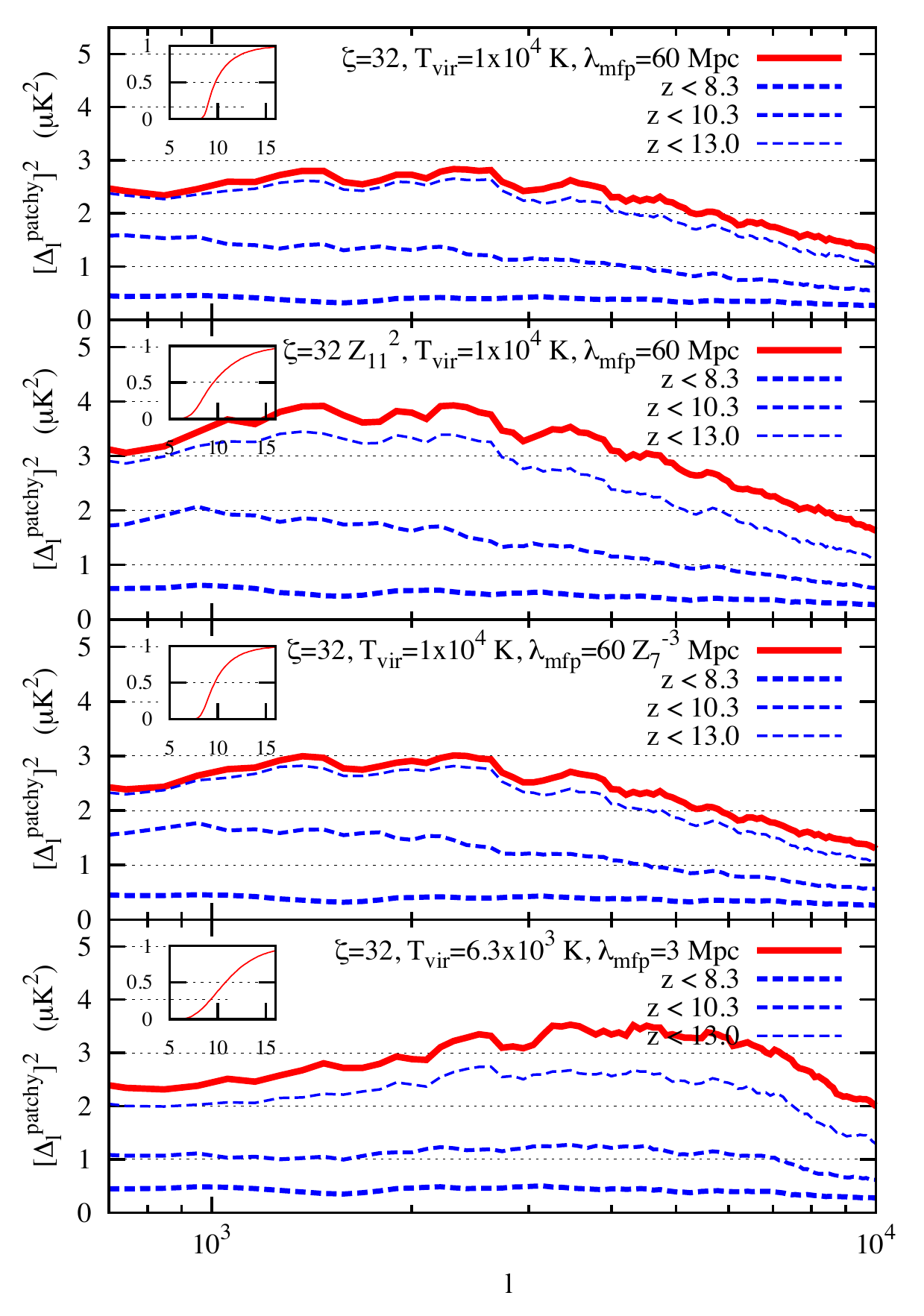}
\caption{
The kSZ dimensionless angular power spectra for the models featured in Fig. \ref{fig:ionization_maps}.  The inset in each panel shows the evolution of $\avenf$ (y-axis) with redshift (x-axis).
\label{fig:ps_breakdown}
}
\vspace{-1\baselineskip}
\end{figure}

\begin{figure}
\vspace{-1\baselineskip}
\includegraphics[width=0.45\textwidth]{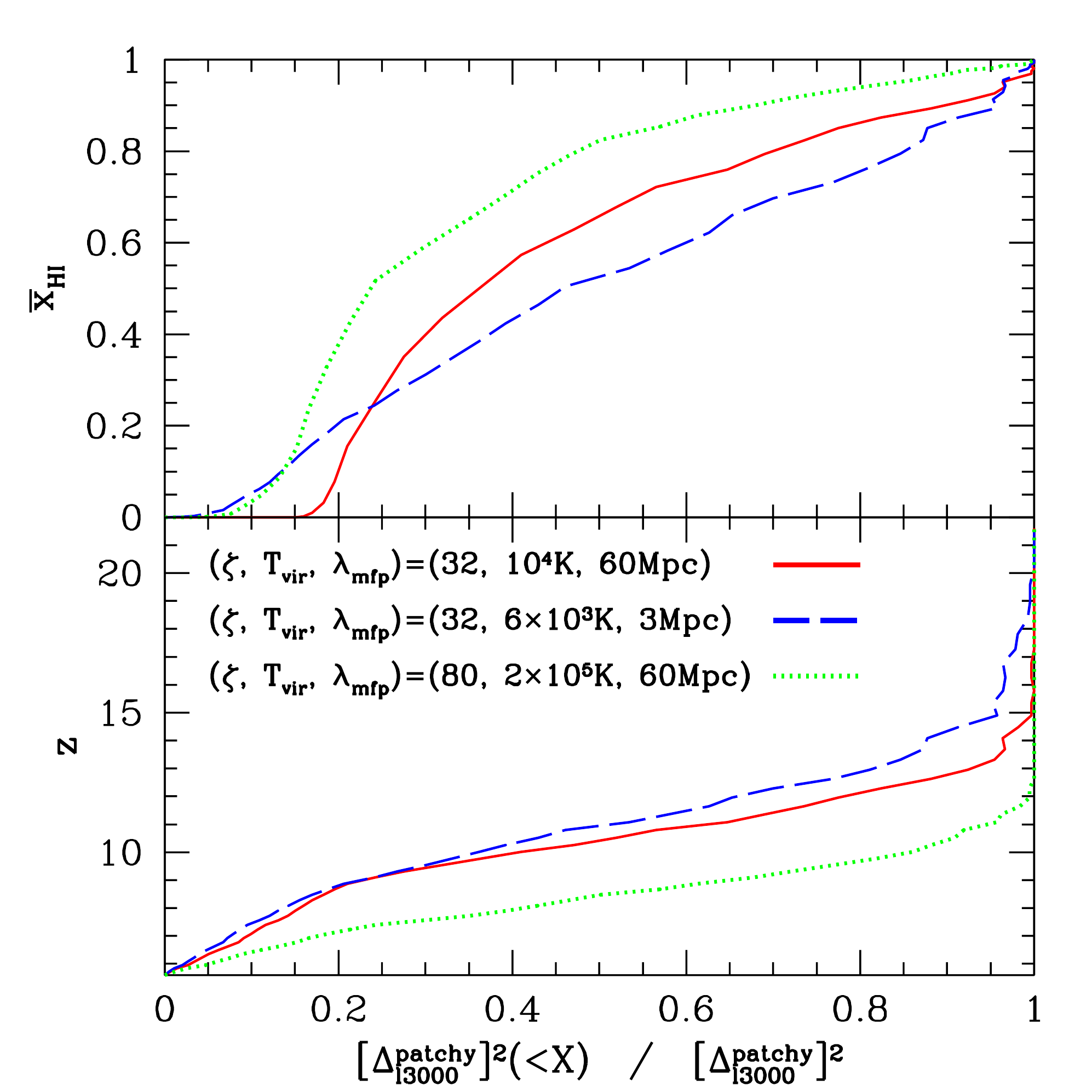}
\caption{
The cumulative fraction of the $z>5.6$ kSZ signal at $l=3000$ sourced by redshifts less than $z$ ({\it bottom panel}) and corresponding global neutral fractions less than $\avenf$ ({\it top panel}). Three model curves are shown corresponding to:
$\reionparams$ = \{32, 10$^4$ K, 60 Mpc\}, \{32, $6\times10^3$ K, 3 Mpc\}, \{80, $2\times10^5$ K, 60 Mpc\}.
\label{fig:kSZ_dist}
}
\vspace{-1\baselineskip}
\end{figure}

In Fig. \ref{fig:ps_breakdown}, we present the kSZ power spectra corresponding to the four models shown in Fig. \ref{fig:ionization_maps}.  The total signal is represented by the solid red curves, with dashed  curves showing the contributions up to the specified redshift.  
The top panels in both figures \ref{fig:ionization_maps} and \ref{fig:ps_breakdown}  correspond to a non-evolving model with $\reionparams$ = \{32, $10^4$ K, 60 Mpc\}.
The second panels down correspond to the same parameters as the top panels, but including redshift evolution of the ionizing efficiency, $\zeta = 32 \,[(1+z)/11]^2$.  By increasing the efficiency in this model at early times ($z>10$) and decreasing it at later times ($z<10$), the ionized structures become more uniform in size, and reionization is extended.  The corresponding power spectra at $l=3000$ (corresponding to $\sim$ 20Mpc structures at high-$z$) is the largest in this model, $\PkSZ=3.3~\muKK$, placing it on the threshold of current measurements.
Such a model could be motivated, for example, by an efficient top-heavy IMF in the early universe, transitioning to an inefficient, feedback-regulated star formation regime at later times. 
We find that models with stronger evolution (e.g., $\zeta = 32\, [(1+z)/11]^3$) are ruled out by present QSO constraints since reionization fails to complete in time (see our discussion of constraints derived from QSO spectra below).  Models in which $\zeta$ decreases with redshift result in a more sudden reionization and a smaller kSZ signal.

The model in the third panels of figures \ref{fig:ionization_maps} and \ref{fig:ps_breakdown}  also assumes a constant $\zeta$ and $\Tvir$ as in the top panel.  However, this model includes evolution in $\mfp = 60 [7/(1+z)]^3$.  This evolution is in rough agreement with extrapolations using the measured trend in the ionizing-photon mean free path from $z\sim$2--4 (e.g., \citealt{PWO09}), though extrapolations into the reionization epoch are extremely uncertain.
The evolving $\mfp$ model results in similar HII bubble morphology as the non-evolving model in the top panel; however, the end stages are extended.

Finally, the bottom panel corresponds to an extreme scenario, where LLSs strongly suppress the growth of HII regions, resulting in a very extended reionization.  The corresponding parameters are $\reionparams$ = \{32, $6.3\times10^3$ K, 3 Mpc\}.   Here ionization structures are smaller compared with other models, shifting the peak in kSZ power to smaller scales.

We also point out that there is evidence in all four panels that the power in the higher $l$-multipoles is sourced from earlier times when the bubbles were smallest (i.e. the blue curves start ``pealing off'' of the red ones from high-$l$ as the redshift upper limit is decreased). As mentioned above, this is due to the fact that the ionization power peaks on increasingly larger scales as reionization progresses and the HII bubbles grow.  Understandably, this trend is least evident in the bottom panel, where HII bubbles are constrained to be small for a longer duration, and the contrast between ionization power on large and small scales is smaller.

As already discussed, our fiducial models do not include redshift evolution of our three astrophysical parameters.  However, it is interesting that for most of the models shown in Fig. \ref{fig:ps_breakdown}, the majority of the kSZ signal is imprinted over a relatively narrow redshift or, more fundamentally, $\avenf$ range.  
We have already noted in \S \ref{sec:kSZ} that doubling the duration of reionization roughly doubles the signal.  More precisely however, the important quantity is not the duration of the entire reionization process, but just of the epoch when ionized structures were of the angular scale of interest.  The timing of this epoch is model-dependent, but we see from Fig. \ref{fig:ps_breakdown} that the ionization morphology had notable power on $l\approx3000$ scales during the early/middle stages of reionization.

To further quantify which epochs source the patchy kSZ signal, in Fig. \ref{fig:kSZ_dist} we plot the fraction of the total $z>5.6$ kSZ signal at $l=3000$ sourced by redshifts less than $z$ ({\it bottom panel}) and global neutral fractions less than $\avenf$ ({\it top panel}).  Three model curves are shown, corresponding to: $\reionparams$ = \{32, 10$^4$ K, 60 Mpc\}, \{32, $6\times10^3$ K, 3 Mpc\}, \{80, $2\times10^5$ K, 60 Mpc\}.  The extremely extended reionization model shown with the blue dashed curve has kSZ power imprinted smoothly and uniformly throughout reionization, i.e. $d\PkSZ/d\avenf \sim 1$.  In this case, as is qualitatively evident from the bottom panel of Fig. \ref{fig:ionization_maps}, the ionization structure retains power on $l=3000$ ($L\approx20$ Mpc) scales throughout reionization.  However, more standard models are not sensitive to the later stages of reionization, when HII regions grow to be larger than $l=3000$.  The ``fiducial'' case shown with the solid red curve only has $\approx20$\% of the $l=3000$ reionization signal imprinted in the last half of reionization, $\avenf<0.5$.\footnote{Note that in this case, $\approx17$\% of the $z>5.6$ signal originates post-reionization, $5.6<z\lsim8.5$, and would be more appropriately classified as the OV component (see the discussion of our conventions at the beginning of \S 2).}  Models with even larger HII regions, such as the one shown with the dashed green curve, are even less sensitive to the later stages of reionization.  In particular, this model has only $\approx16$\% of the $l=3000$ reionization signal imprinted in the last half of reionization.
 The fact that the kSZ signal is sensitive mostly to the first half of reionization suggests that our astrophysical parameters can be interpreted as the average values during this epoch.

\subsection{Signal at $l=3000$}

For the remainder of the paper, we focus our attention of the power at $l\approx3000$, using the statistic $\PkSZ$ $=3000^2 \, C^{\rm patchy}_{l=3000} /(2 \pi)$.
As discussed in the introduction, these multipoles  are the most observable with upcoming and future small-scale CMB observations.  
  At $z\gsim 5$, $l=3000$ roughly corresponds to a physical distance of $\sim \eta 2 \pi/l =  20$ Mpc.  This suggests that $\PkSZ$ depends on the timing and duration of ionization structure of comparable scale.

\begin{figure*}
\vspace{+0\baselineskip}
{
\includegraphics[width=0.45\textwidth]{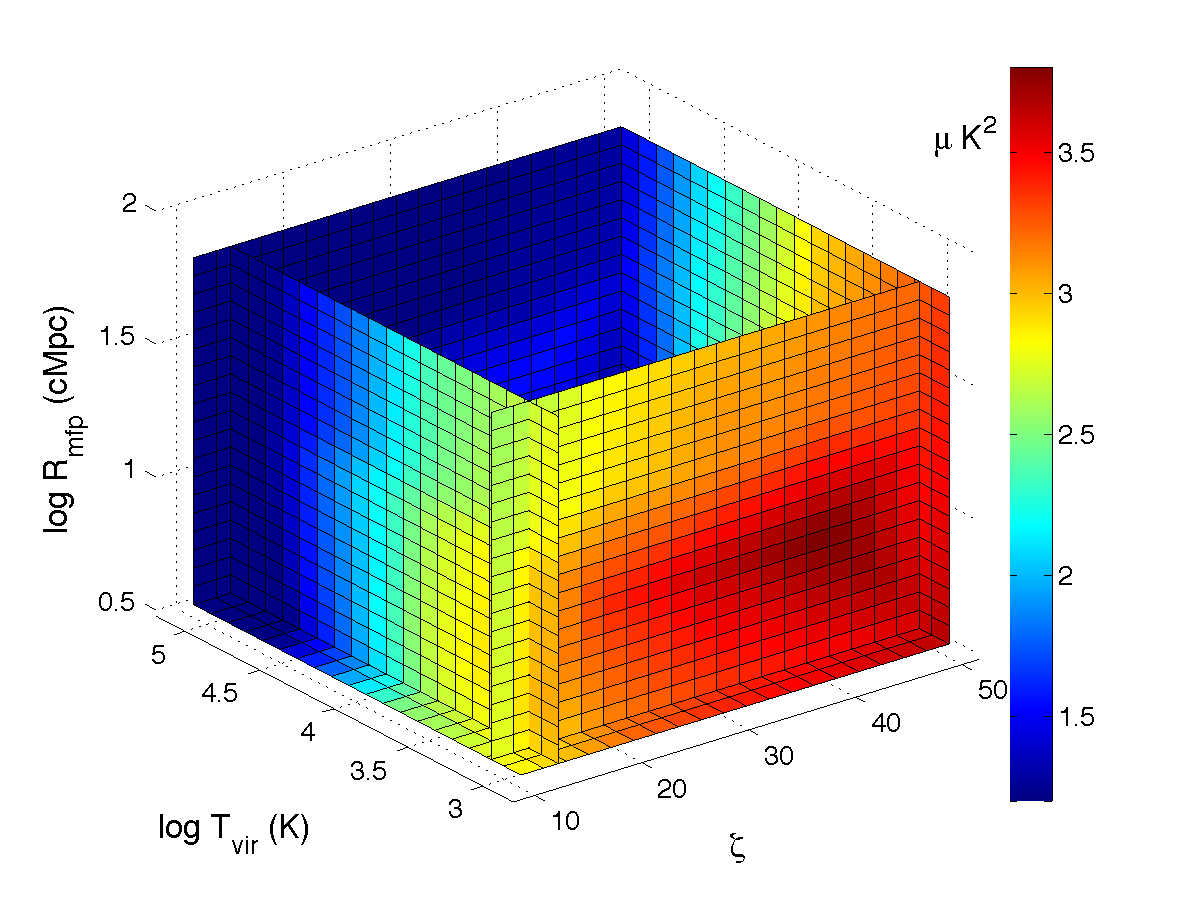}
\includegraphics[width=0.45\textwidth]{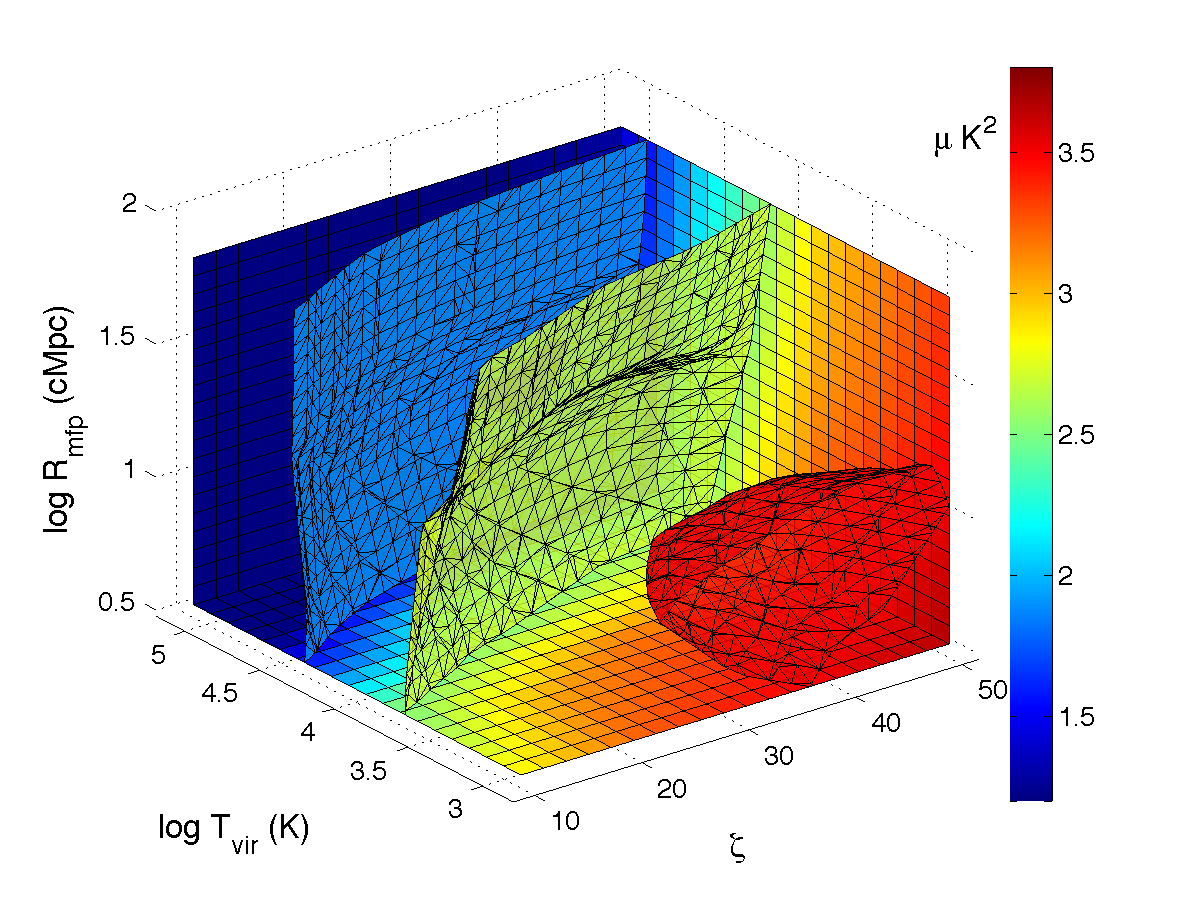}
}
\caption{$\PkSZ$ at $l=3000$ from inhomogeneous reionization, as a function of our three astrophysical parameters. {\it Left panel:} Slices through the 3D parameter space are shown at $\zeta=$ 15 and 52; $\Tvir= 10^{3.2}$ and $10^{5.2}$ K; and $\mfp=$ 3 Mpc. {\it Right panel:} Cuts through the parameter space are shown with $\PkSZ=$ 3.5, 2.5, 1.5$\muKK$.
\label{fig:3d_power_boxes}
}
\vspace{-1\baselineskip}
\end{figure*}

In Fig. \ref{fig:3d_power_boxes}, we plot $\PkSZ$ as a function of $\reionparams$.  Values of $\PkSZ$ were interpolated from our 103 \cmfast\ samples using tesselation-based linear interpolation.  Slices through the 3D parameter space are shown at $\zeta=$ 15, 52; $\Tvir= 10^{3.2}, 10^{5.2}$ K; and $\mfp=$ 3 Mpc ({\it left panel}). The $\PkSZ=$ 3.5, 2.5, 1.5$\muKK$ contours are plotted in the right panel.

In our parameter space the signal ranges from $\PkSZ\approx$ 1 -- 4 $\muKK$.  
  The strongest signal in Fig. \ref{fig:3d_power_boxes} corresponds to the regime with early, extended reionization scenarios.
  In other words, high values of $\PkSZ$ result when reionization is driven by very small galaxies (small $\Tvir$) whose contribution is significant (large $\zeta$) so as to start reionization early, and with relatively abundant LLSs (low $\mfp$), so as to extend the duration of the epochs where HII regions are comparable in size to $l\approx3000$.  These scenarios would be the easiest to observe, and the first to be ruled out with an upper limit on $\PkSZ$.  

There are a couple of interesting non-monotonic trends to note in Fig. \ref{fig:3d_power_boxes}.  The peak in power occurs at $\reionparams \sim$ \{40, 10$^3$ K, 10 Mpc\}.  As the ionizing efficiency from this peak is increased, the kSZ power falls somewhat because this change results in reionization occurring more rapidly.  Lower efficiencies delay reionization to later epochs, also resulting in a slightly weaker kSZ signal.

Additionally, the smaller the mean free path generally the stronger the signal, since they produce more extended reionization histories.  However, the $l = 3000$ power is a maximum for models with $\mfp \approx 10$ Mpc because this $\mfp$ is roughly the scale probed at this multipole  (see left panel Fig. \ref{fig:3d_power_boxes}), whereas smaller $\mfp$ values suppress structure on this scale.   

\subsubsection{Including constraints from WMAP and QSOs}
\label{sec:constraints}

\begin{figure*}
\includegraphics[width=0.7\textwidth]{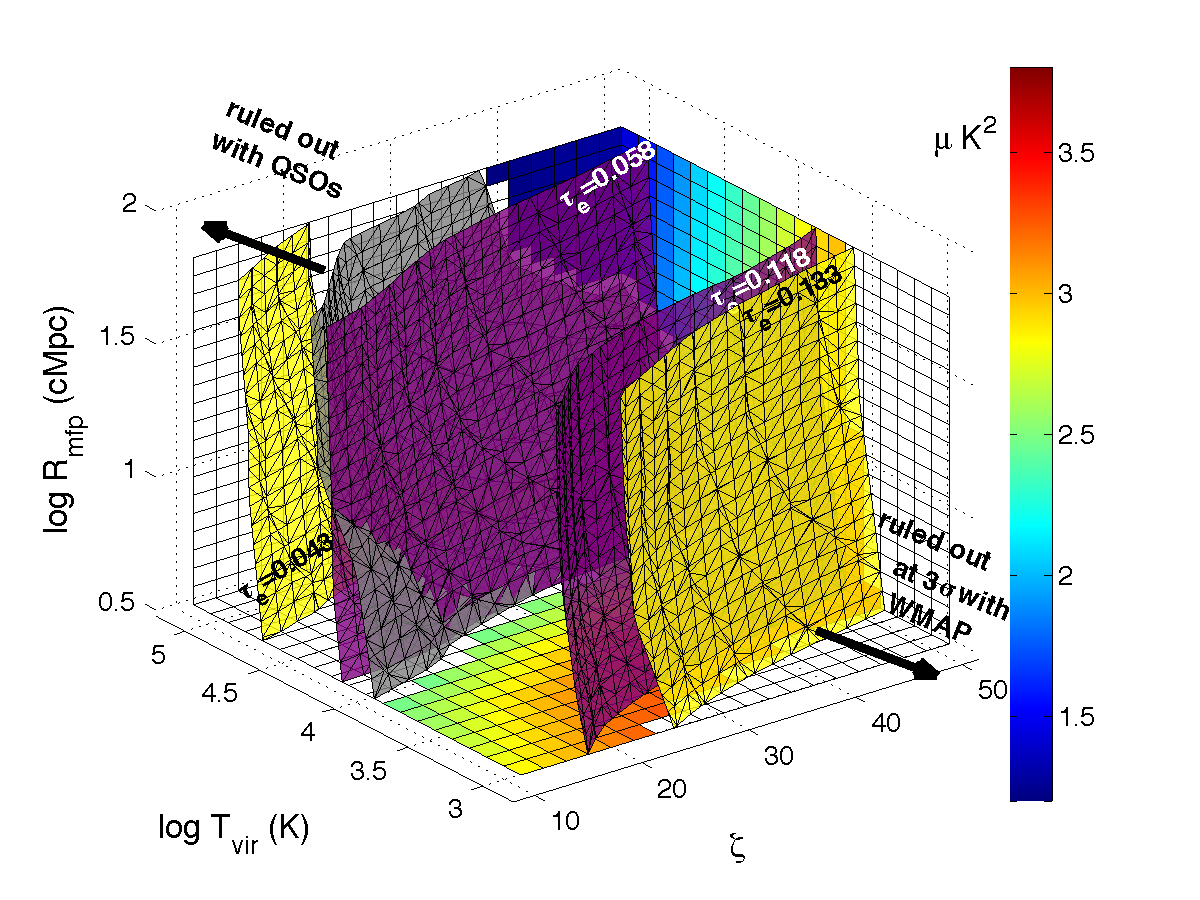}
\caption{
3D parameter space demarcating regions that are inconsistent with current QSO and WMAP7 $\tau_e$ constraints (see text).  The gray contour corresponds to models which have $\avenf=0.25$ at $z=5.6$.  The purple (yellow) contours demarcate the 2 (3) $\sigma$ WMAP7 limits on $\tau_e$.
\label{fig:constraints}
}
\vspace{-1\baselineskip}
\end{figure*}

Other probes of the reionization epoch can compliment the kSZ signal, and restrict the allowed parameter space.  Our two most robust constraints at present come from the Lyman-series forests detected in the spectra of high-redshift QSOs and from the measurement of $\tau_e$ by the {\it WMAP} satellite.

The Lyman-series forests in high-redshift QSOs, originally detected in the Sloan Digital Sky Survey (SDSS), have been used in recent years to claim that reionization must have been over by $z\sim6$ (e.g., \citealt{Fan06}).  Photons emitted by a QSO blueward of the Ly$\alpha$ line can redshift into the Lyman resonances along the LOS and be absorbed by atomic hydrogen (although complete absorption does not necessarily indicate a neutral region since this probe saturates easily; \citealt{GP65}).
  A model-independent, maximally conservative upper limit on $\avenf$ can be obtained just from the fraction of pixels in the Lyman-$\alpha$ and Lyman-$\beta$ forest which are jointly dark \citep{Mesinger10}, yielding $\avenf\lsim0.25$ at $z=5.6$ (\citealt{MMF11}; McGreer et al., in preparation).  
Fig. \ref{fig:constraints} demarcates in gray the surface corresponding to models which have $\avenf=0.25$ at $z=5.6$.  The parameter space on the other side of that surface is robustly ruled out with QSO measurements, and we leave it blank in this and future plots.  As expected, QSO constraints rule out models where the sources reside in very rare, massive halos such that reionization is late or those with a significant $\avenf$ ``tail'' to lower redshifts owing to a small mean free path.  

\begin{figure*}
\includegraphics[width=0.45\textwidth]{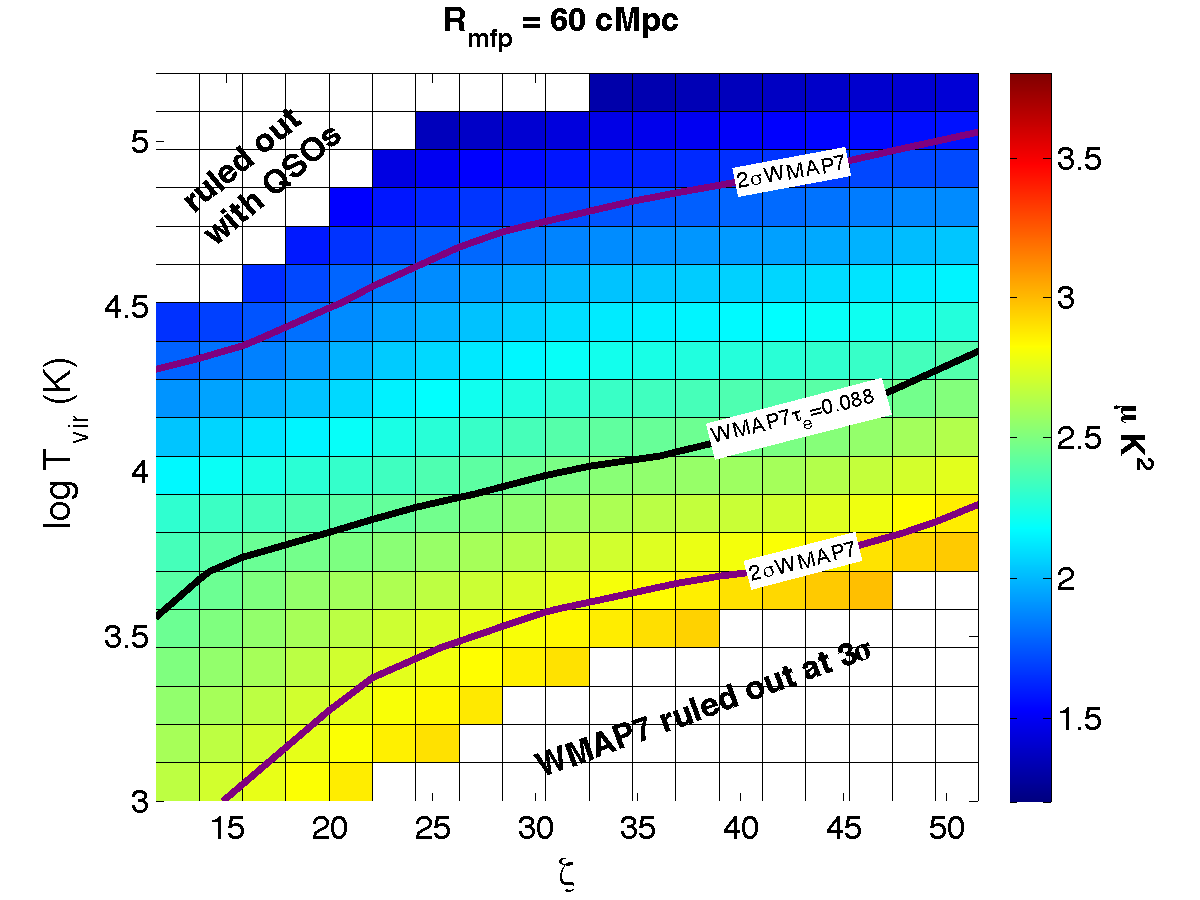}
\includegraphics[width=0.45\textwidth]{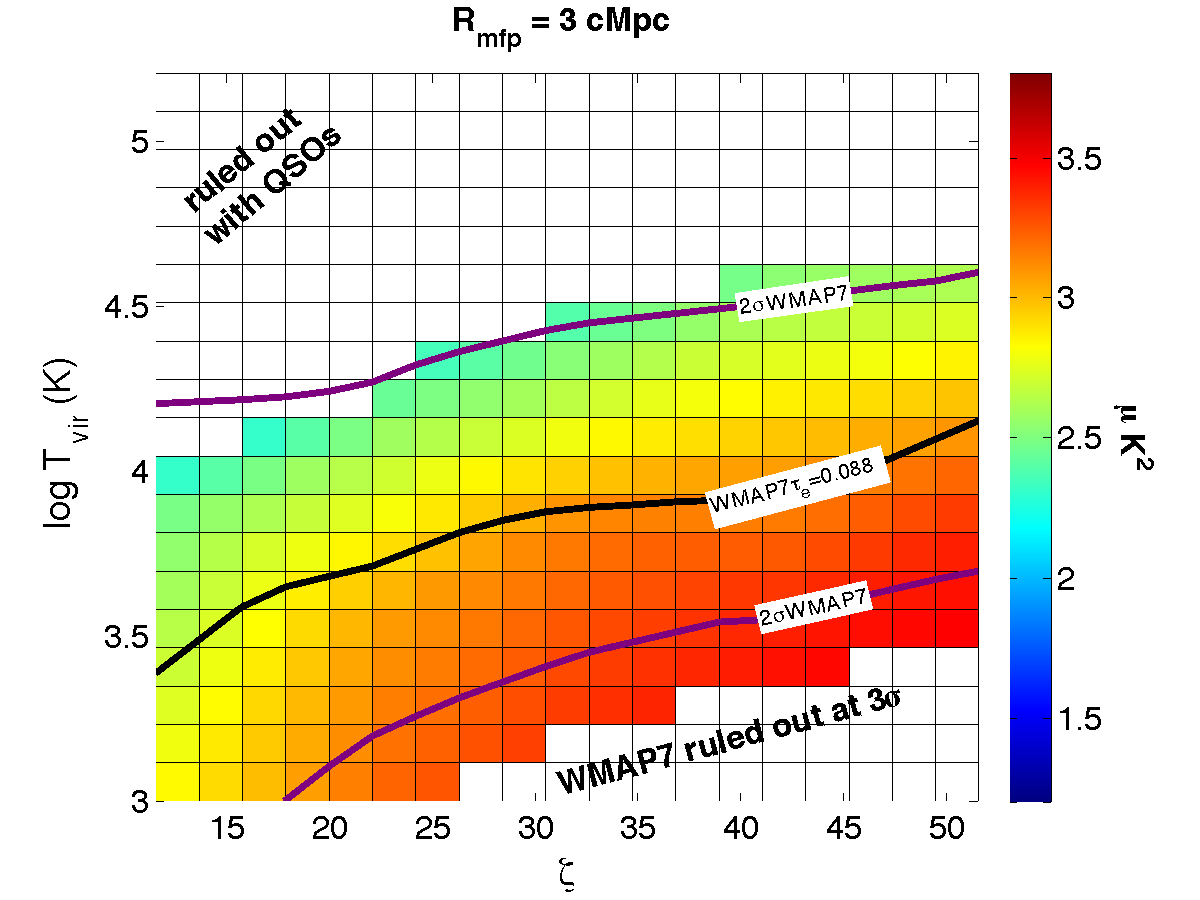}
\includegraphics[width=0.45\textwidth]{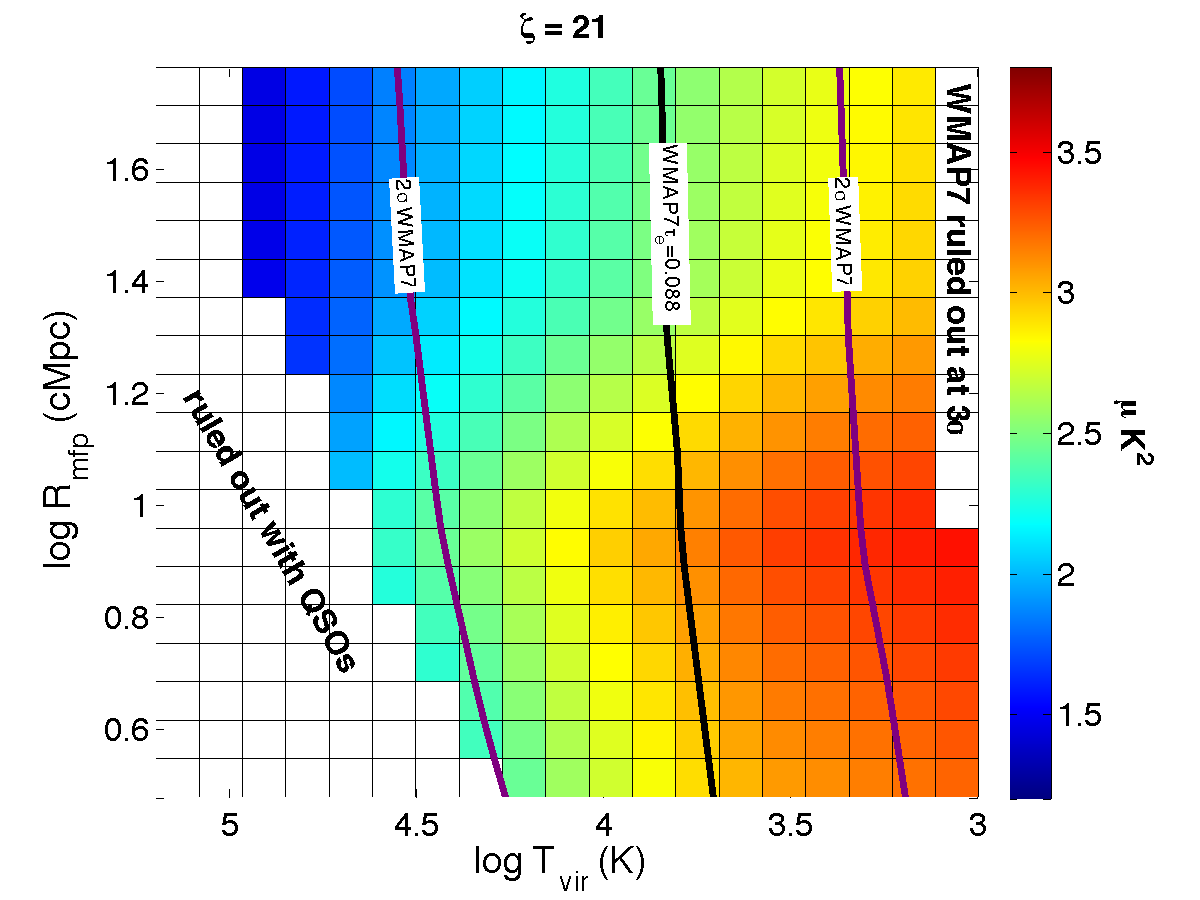}
\includegraphics[width=0.45\textwidth]{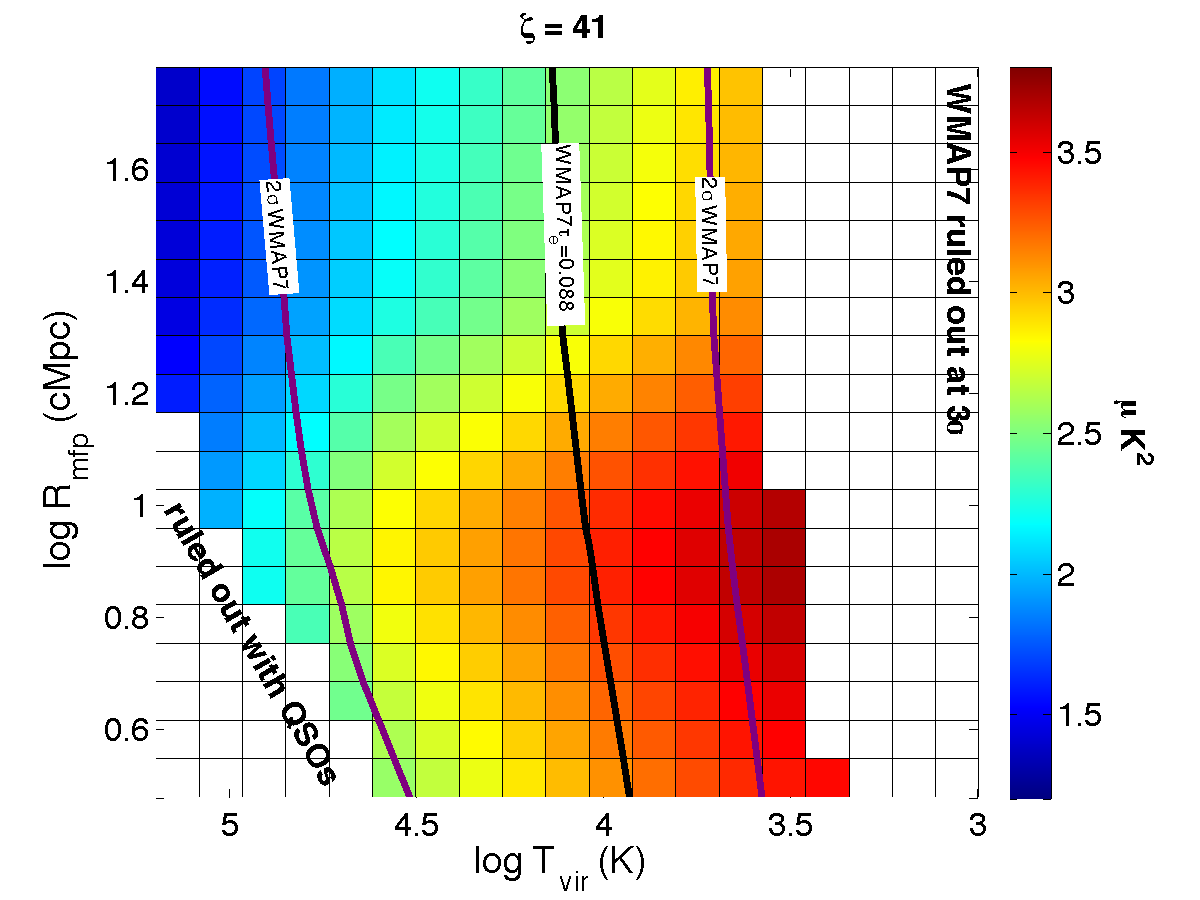}
\includegraphics[width=0.45\textwidth]{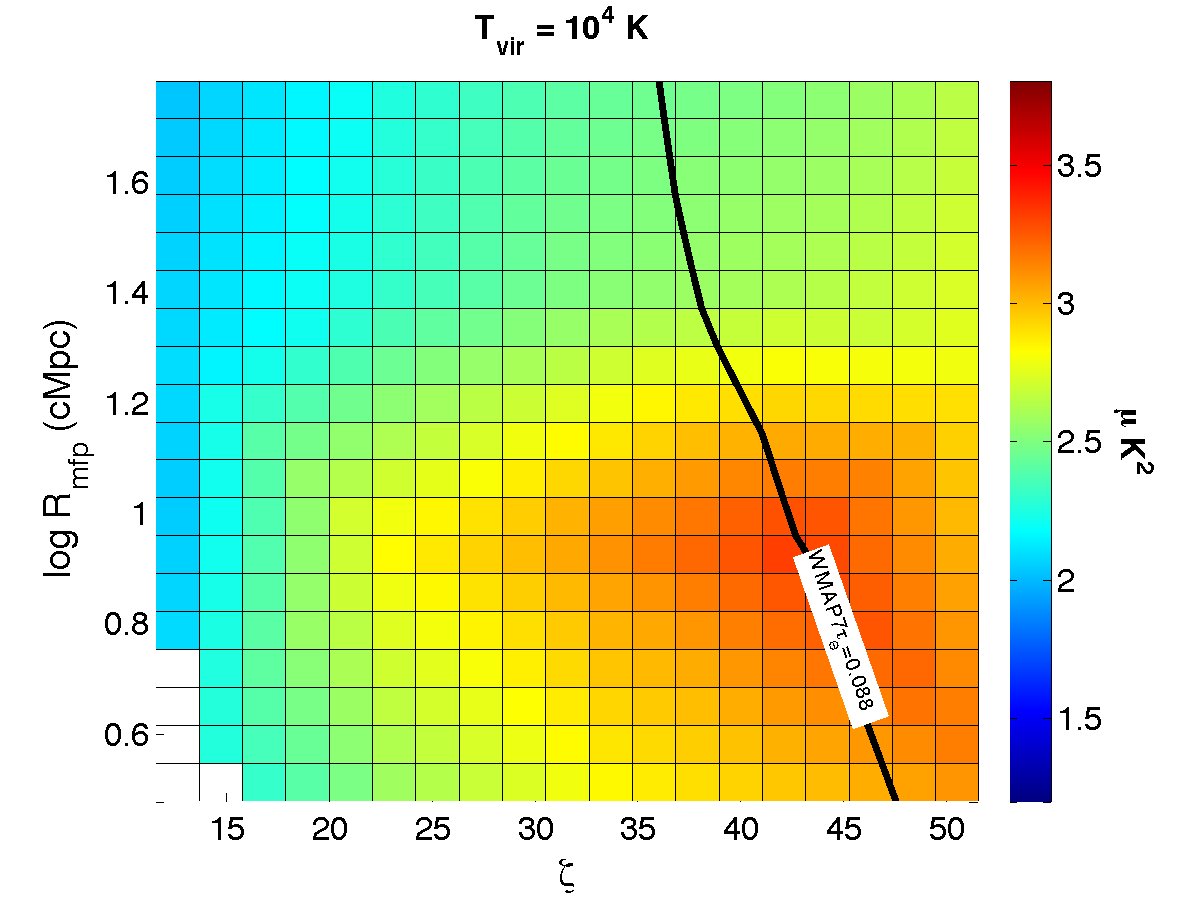}
\includegraphics[width=0.45\textwidth]{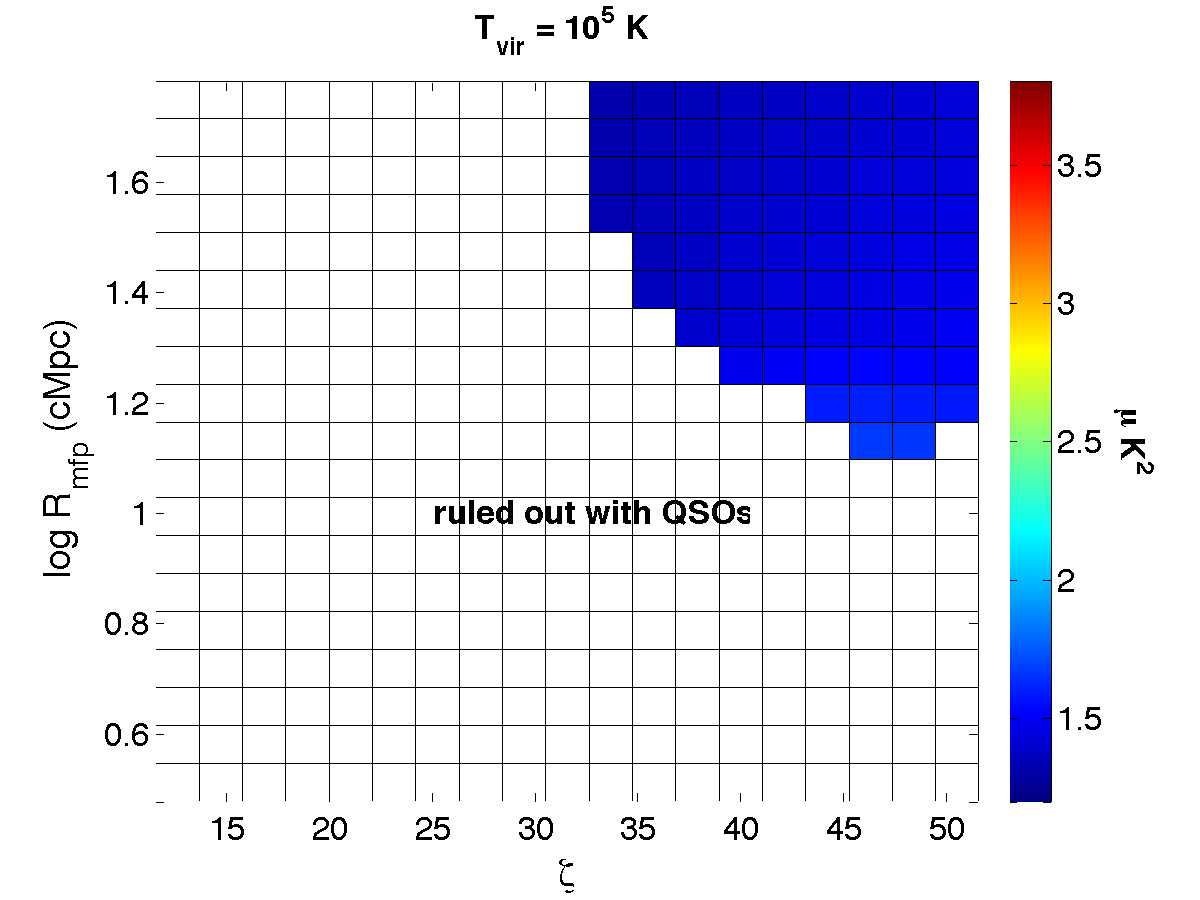}
\caption{
 Slices of $\PkSZ$ through our astrophysical parameter space.  {\it Top:} $\mfp=$ 60 and 3 Mpc; {\it middle:}  $\zeta$ = 21 and 41; {\it bottom:} $\Tvir=$ 10$^4$ and 10$^5$ K.  Also shown are contours at fixed $\tau_e$, corresponding to the prefereed WMAP7 $\tau_e$ value and its 2$\sigma$ limits. 
\label{fig:slices}
}
\vspace{-1\baselineskip}
\end{figure*}

The other significant constraint on reionization comes from the measurement of the Thompson scattering optical depth to CMB photons.  The seven-year {\it WMAP} (WMAP7) estimate is $\tau_e=0.088\pm0.015$ \citep{Komatsu11}.   Fig. \ref{fig:constraints} demarcates the 2$\sigma$ WMAP7 limits on $\tau_e$ with purple surfaces and the 3$\sigma$ limits with yellow surfaces.  We consider the portion of our parameter space not enclosed by the 3$\sigma$ contours to be robustly ruled out, and we leave it blank in this and future plots. The WMAP7 $\tau_e$ limits mainly serve to rule out very early reionization scenarios. These are the models where reionization is driven by very small galaxies.  Fig. \ref{fig:constraints} and Fig. \ref{fig:slices} demonstrate that {\it an upper limit of $\PkSZ\lsim3.5$ improves on existing 3$\sigma$ WMAP7 constraints.}

Fig. \ref{fig:slices} shows additional slices through the reionization parameter space, to focus on the trends noted above.  The top row shows $\PkSZ$ for cuts at $\mfp=$ 60 and 3 Mpc.  Consistent with our previous discussion, $\PkSZ$ increases with decreasing $\mfp$ (until very low values) as reionization becomes more prolonged.  

The middle row of panels in Fig. \ref{fig:slices} shows $\PkSZ$ at $\zeta$ = 21 (left panel) and 41 (right panel).  Contours of constant kSZ power shift to more massive sources as the ionizing efficiency is increased (i.e. the kSZ signal increases with $\zeta$, since reionization occurs earlier).  The non-monotonic trend of $\PkSZ$ with $\mfp$ is clearly evident in these panels. The signal is not sensitive to the value of the mean free path when $\mfp \gsim 15$ Mpc, since these values are large enough that they do not impact the reionization morphology and the morphology is driven just by the clustering of the sources.  

The bottom row of Fig. \ref{fig:slices} shows slices at $\Tvir= 10^4$ and 10$^5$ K.  A similar structure of the signal is seen as in the top panels, with the signal now decreasing with $\Tvir$.  High redshift QSOs already rule out most of our parameter space at $\Tvir=10^5$ K.  
Although slightly fainter than current direct detection limits (from Hubble/WFC3, see e.g., \citealt{SFD11, FDO11}), these sources reside in halos massive enough to be rare in the early Universe.  
  If such relatively massive sources drove reionization, it would have to occur at late epochs and occur very rapidly not to violate existing constraints, resulting in a weak patchy kSZ signal, $\PkSZ\approx 1 \muKK$.  Even so, these scenarios are still ruled out by WMAP7 at $>2\sigma$.

When viewed at constant $\tau_e$, the reionization signal varies by $d \PkSZ \approx 1 \muKK$. If the signal can be determined to such accuracy (see \S \ref{sec:detect}), the kSZ would provide {\it complimentary} information to the mean optical depth statistic.  This complementary information can be used to determine the duration of reionization, $\Delta z_{\rm re}$, as we shall see below.  In our parameter space, the duration of reionization is most sensitive to the mean free path in the range $\mfp\lsim15$ Mpc. Nevertheless, smaller differences of $d \PkSZ \sim 0.5 \muKK$ can be obtained by varying just $\Tvir$ and $\zeta$ at fixed $\tau_e$ (see the top row of Fig. \ref{fig:slices}).

\subsubsection{Redshift and duration of reionization}
\label{sec:zre_delz}

Here we cast our patchy kSZ signals in terms of two empirical parameters, which can be easier to interpret than our astrophysical ones:
\begin{packed_enum}
\item $\zre$: {\it defined as the redshift at which $\avenf=0.5$,}\\
\item $\delz$: {\it defined as the redshift interval from $\avenf=0.75$ to $\avenf=0.25$.}
\end{packed_enum}
Note that by definition $\Delta \zre$ is a change in the HI fraction by $\Delta \avenf$=0.5.  The entire process of reionization could be considerably longer owing to extended early and late epochs (e.g., \citealt{Mesinger10proceedings}), but the kSZ signal at $l\sim3000$ is less sensitive to these stages as demonstrated above.

\begin{figure}
\includegraphics[width=0.45\textwidth]{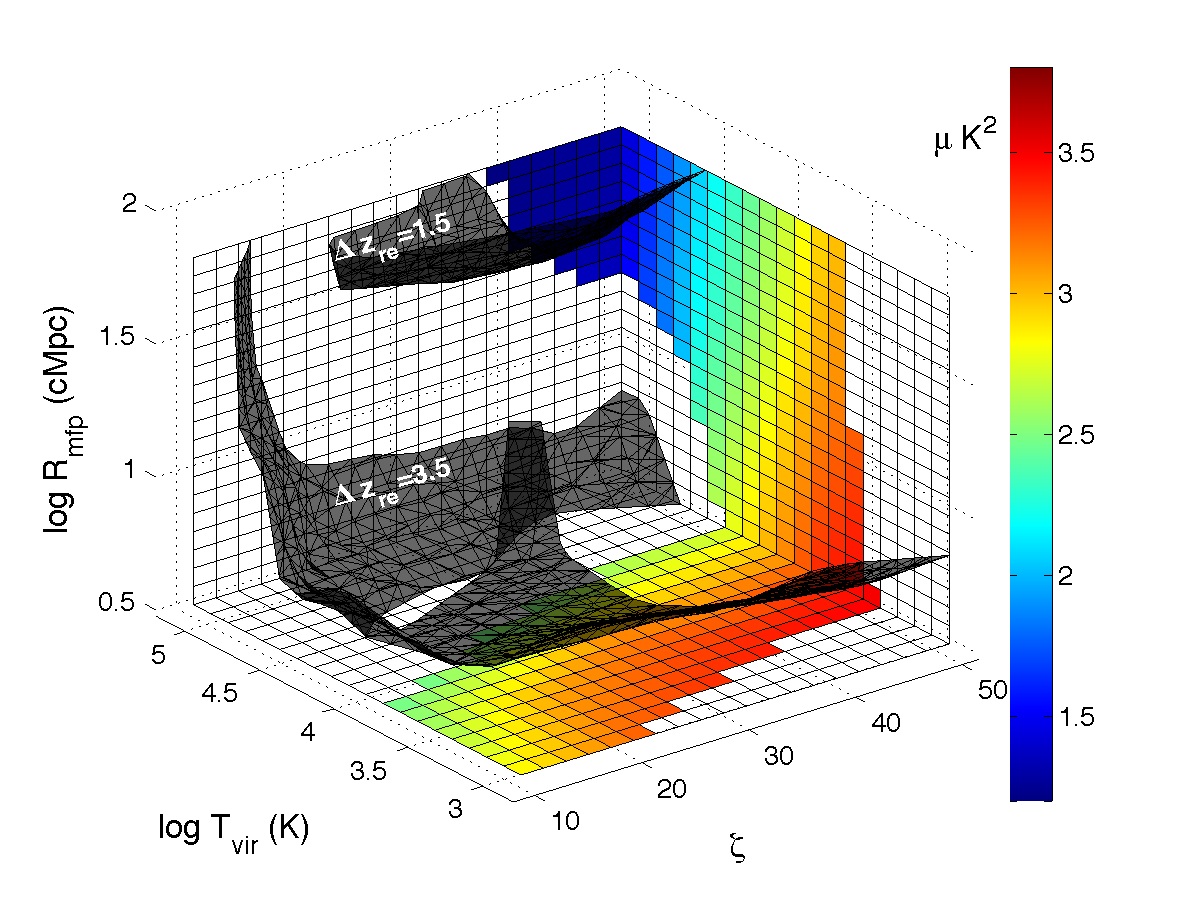}
\caption{
Contours of $\delz=$ 1.5 and 3.5 through our astrophysical parameter space ({\it upper} and {\it lower} contours, respectively).  Note that  $\delz$ is defined as the interval for which $0.25 < \avenf < 0.75$, which can be significantly shorter than the duration over a larger range in $\avenf$. 
\label{fig:3d_delz}
}
\vspace{-1\baselineskip}
\end{figure}

Fig. \ref{fig:3d_delz} shows contours of fixed $\delz=$ 1.5 and 3.5 through our astrophysical parameter space.  Understandably, rapid reionization scenarios are located in the top, back right corner of this figure: bright sources hosted by massive halos with relatively few LLSs to impede the progress of reionization.
  The contrary is true for slow reionization models.  For clarity of presentation, surfaces of constant $\zre$ are not shown; however, they are very similar to the constant $\tau_e$ contours shown in the previous figures. 

\begin{figure}
\includegraphics[width=0.45\textwidth]{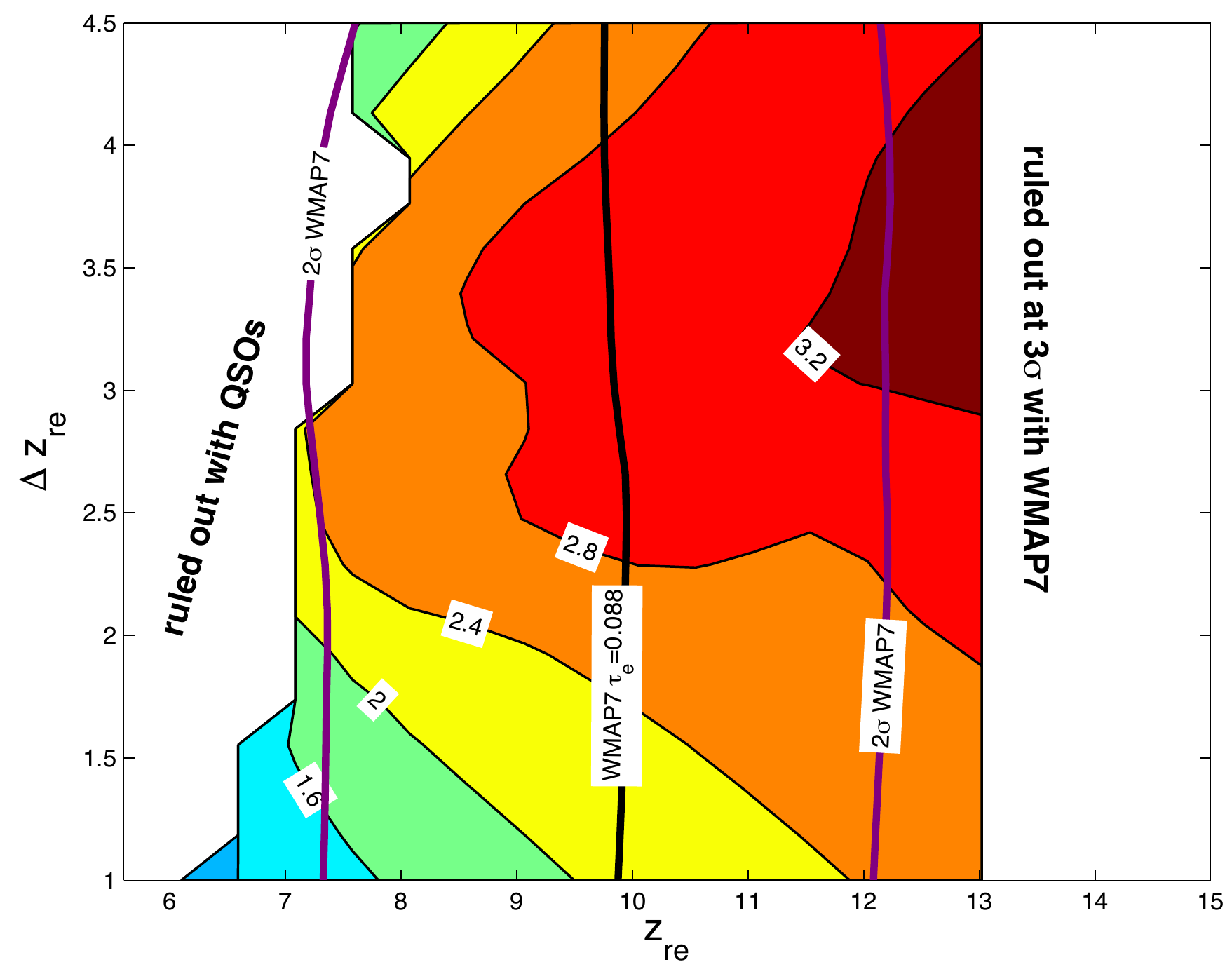}
\caption{
Contours of $\PkSZ$ as a function of $\zre$ and $\delz$.
\label{fig:zre_delz}
}
\vspace{-1\baselineskip}
\end{figure}

Fig. \ref{fig:zre_delz} shows contours of $\PkSZ$ as a function of $\zre$ and $\delz$. Constant $\tau_e$ lines are also depicted.  These almost correspond to lines of constant $\zre$, since the reionization histories of our models are relatively symmetric around $\zre$.\footnote{Note that the curving of the lower 2$\sigma$ limit on $\tau_e$ as one approaches high values of $\delz$ results from the fact that these extreme scenarios (corresponding to late-appearing sources and small mean free paths), have an asymmetric extended tail to low $\avenf$ (see Fig. \ref{fig:ps_params}).  In these models the small mean free path to ionizing photons creates an effective horizon thus delaying the completion of the final (i.e. overlap) stages of reionization, as discussed above \citep{FM09, AA10}.}

A constraint of $\PkSZ < 3.2 \muKK$ would improve on the WMAP7 2$\sigma$ limit on $\zre$ in our models.  A detection of $\PkSZ$ combined with the measurement of $\tau_e$ could place complimentary constraints on both the epoch and duration of reionization.  However, as we have already seen, a measurement of $\PkSZ$ and $\tau_e$ in general does not uniquely determine $\delz$.  For example, $\tau_e=0.088$ and $\PkSZ=2.8 \muKK$ allow both $\delz=$ 2.3 and 4.0.  This is primarily due to the fact that the peak of the kSZ power shifts in angular scale.

\begin{figure}
\includegraphics[width=0.45\textwidth]{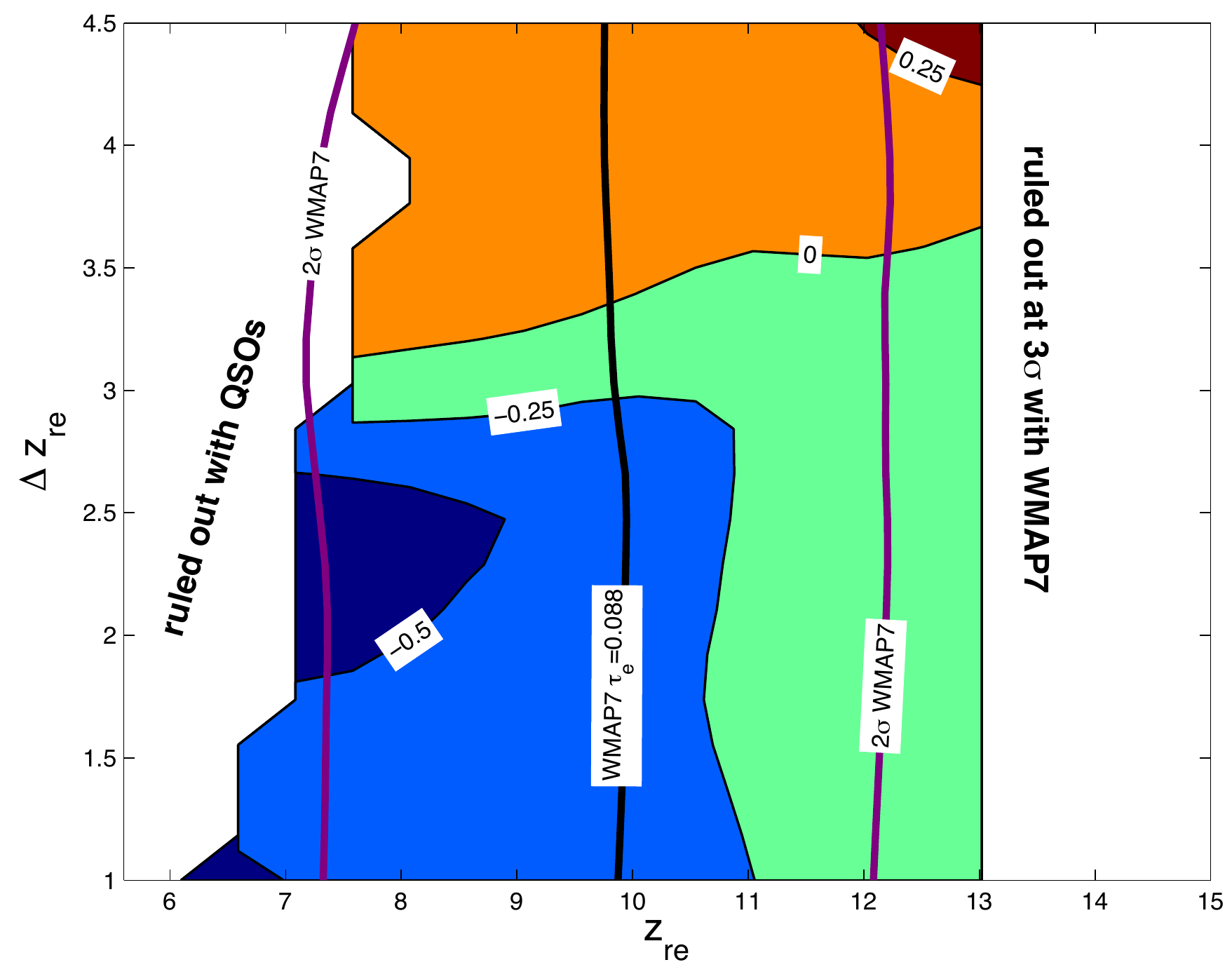}
\caption{
Contours of the logarithmic slope of the power spectrum around $l\sim3000$.  Specifically, the slope is defined as 2\,log[$\Delta^{\rm patchy}_{l3300}/\Delta^{\rm patchy}_{l2700}$] / log[3300/2700].
\label{fig:slope_zre_delz}
}
\vspace{-1\baselineskip}
\end{figure}

The {\it slope} of the patchy kSZ power at $l=3000$ can break this degeneracy.  The slope requires sensitivities that are a few times improved over ongoing measurements that seek to detect the kSZ.  This logarithmic slope is plotted in Fig. \ref{fig:slope_zre_delz}.\footnote{This figure was created with a wider spectral smoothing kernel ($\Delta l=600$, instead of $\Delta l=300$ used throughout the rest of the paper).  This was done to smooth over small spectral features present in our realization.  We find that, unlike $\PkSZ$,  the spectral slope as defined above is sensitive to such cosmic variance effects.}
  From this, we see that the longer/earlier reionization scenarios have a positive slope, meaning that the peak in kSZ power occurs on scales smaller than $l=3000$, which correspond to earlier stages of reionization (see the bottom panel of Fig. \ref{fig:ps_breakdown} and associated discussion).
  A detection of the sign of this slope, combined with $\tau_e$ and $\PkSZ$, should robustly determine $\zre$ and $\delz$.  Furthermore, we find that positive values of the slope come exclusively from models with low values of $\mfp \lsim 10$ Mpc.  Therefore detecting the sign of the slope can constrain the abundance of absorption systems in the early Universe, at redshifts far beyond those of present estimates based on QSO spectra.


\section{Comparison with Prior Studies}
\label{sec:prior}

Several studies have investigated the kSZ signal from inhomogeneous reionization using either radiative transfer simulations of reionization or  semi-numeric models of this process\footnote{Most previous studies used a cosmology with $\sigma_8 = 0.9$, which will result in a slightly higher amplitude since the patchy kSZ is proportional to  $v^2 \propto \sigma_8^2$.  $\sigma_8$ primarily affects the timing of reionization, and not the HII morphology \citep{McQuinn07}.}:
\begin{itemize}
\item \citet{McQuinn05} investigated the kSZ from reionization using an analytic framework also based on the \citet{FZH04} model, by varying the parameter $\zeta$; \citet{FMH05} generalized their calculation to include models where the ionizing efficiency depends on halo mass.  This analytic method is more approximate than the calculations presented here.  However, values of $\PkSZ$ found in \citet{McQuinn05} are similar to what we find in our analogous models (although, their functional form for $\PkSZ$ is slightly more peaked).  However, the mass-dependent $\zeta$-models in \citet{FMH05} exhibit a stronger {\it l}-dependence than we find in our $T_{\rm vir}$ models, which should mimic the same effect.  These differences likely owe to the approximations made in their analytic derivations.
\item \citet{Zahn05} calculated the kSZ from two reionization models with a fixed and temporally evolving $\zeta$ parametrization using the \citet{Zahn07} semi-numeric implementation of reionization on top of a $100~$Mpc/h SPH simulation.  This box size is insufficient to capture the largest-scale velocity modes and, thus, should underestimate the kSZ signal. Furthermore, it is difficult to compare our calculations with theirs owing to their different reionization histories.\footnote{The \citet{Zahn07} constant $\zeta$ model has a smaller $\delz$ than ours owing to their implementation.}  Nevertheless, we note that their results qualitatively agreed with those of \citet{McQuinn05} and, by extension, those here.
\item \citet{Iliev07kSZ} used full radiative transfer simulations of reionization to calculate the kSZ from a $100~$Mpc/h simulation supplemented with a scheme to account for missing large-scale velocity flows.  Their power spectra peak at ${\it l} \approx 3000$, similar to our constant $\zeta$ models, but have $2-3$ times higher amplitudes.  The likely cause of this discrepancy is the presence of a more uniform bubble size in the \citet{Iliev07kSZ} models.  This is in contrast with our models and also with other radiative transfer simulations of reionization (e.g., \citealt{McQuinn07, TC07}).
\item When our work was nearing completion, a similar study was made public, focusing on reionization timing and duration constraints implied by the two year SPT data \citep{Zahn12}.  They used a reionization model similar to ours, and although detailed comparisons are difficult, their power spectra roughly agree with ours in both shape and amplitude.  However, they do not explore the same ($\zeta, \Tvir, \mfp$) parameter space we do.  Instead, they only focus on the efficiency parameter, $\zeta$, encoding the sharpness of reionization through an ``acceleration'' parameter.  This effectively allows the ionized fraction to rise much more steeply, $\bar{x}_i\propto (\zeta_0 f_{\rm coll})^\beta$, where they vary $\zeta_0$ and $\beta$ in order to sample a wide range of reionization histories.  Because of this exponential dependence on the collapse fraction, they are able to generate much sharper reionization histories, useful for their Markov chain Monte Carlo (MCMC) analysis.  For example, halving the duration of reionization (and roughly the kSZ power) in a fiducial $\Tvir\sim10^4$K, $\zeta_0\sim20$ model requires $\beta \sim 2.5$.  Although such a parametrization allows for a larger variety of kSZ signals, which can fit the aggressive SPT bounds (see below),  we caution that it is not physically motivated.  Any reasonable feedback mechanism delays reionization instead of accelerating it (e.g., \citealt{TW96, SH03, PS09}). \footnote{The very first, molecularly cooled galaxies could experience positive/accelerating feedback due to the enhanced free-electron fraction from ionizing photons (e.g., \citealt{OH02}) or hydrodynamical shocks \citep{SK87}, which catalyzes the formation of molecular hydrogen.  Such feedback would be mild, confined to fossil HII regions and edges of ionization fronts (e.g., \citealt{RGS02b, KM05}), and short-lived (\citealt{HAR00, MBH06}).}  Our work here is complementary to theirs, as we predict the kSZ signal for a wide range of realistic reionization models. 
We also provide physical insights into the signal, for example showing that the shape of the patchy kSZ signal can be significantly affected by $\mfp$; thus $\PkSZ$ and $\tau_e$ do not uniquely determine the redshift and duration of reionization.  We also demonstrate that the bulk of the patchy signal for most models is sourced by the early-to-middle stages of reionization.  Therefore any kSZ-based inferences about the later stages of reionization are indirect and very model dependent. 
\end{itemize}

\section{Context with Recent and Forthcoming Observations}
\label{sec:detect}

Recently, \citet{Reichardt11} placed a limit of $\Ptot< 2.8~\muKK$ at 95\% C.L. using two year SPT measurements.  This aggressive bound is valid under the assumption that there is no CIB-tSZ correlation.  Allowing for such a correlation, the constraint degrades to $\Ptot \lsim 6~\muKK$ at 95\% C.L..  This is also in agreement with the 1yr data from the ACT, which implies $\Ptot =  6.8\pm 2.9~\muKK$ \citep{Dunkley11}. 

Despite the wide parameter space of models we explore, we predict a fairly narrow range for the patchy reionization component of the kSZ power, $\PkSZ \approx$ 1.5--3.5 $\muKK$.  When combined with estimates for the homogeneous ionization (OV) component in \S \ref{sec:OV} (see also \citealt{SRN11}), $\POV \approx$ 2--3 $\muKK$, this work implies that the total kSZ power is $\Ptot \approx$ 3.5--6.5 $\muKK$.  Therefore the recent conservative SPT bound is on the cusp of constraining viable reionization scenarios.

\begin{figure}
\includegraphics[width=0.45\textwidth]{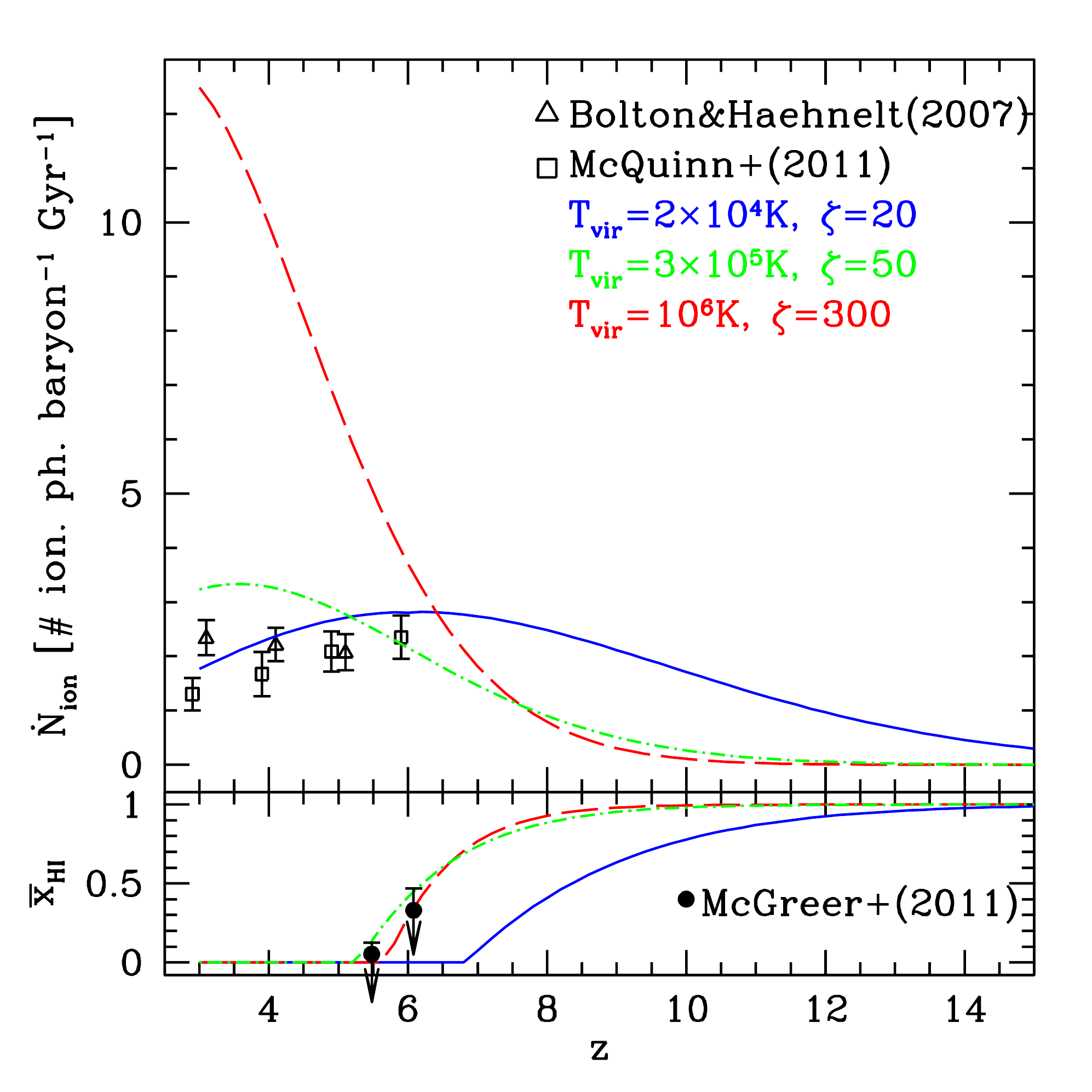}
\caption{
The redshift evolution of the ionizing photon emissivity ({\it top panel}) and $\avenf$ ({\it bottom panel}).  Emissivity constraints inferred from the \lya\ forest are shown as squares \citep{MOF11} and triangles \citep{BH07} (offset by $\Delta z=0.1$ to improve legibility); the maximally conservative neutral fraction constraints from the dark fraction in the \lya\ and Ly$\beta$ forests \citep{MMF11} are shown in the bottom panel.  We also show the analytic predictions, $\dot{N}_{\rm ion} = \zeta (1+n_{\rm rec})df_{\rm coll}/dt$, for three models, assuming $n_{\rm rec}=0$ and $\mfp=\infty$.  The solid blue curve corresponds to ``standard'' parameter choices of $\Tvir=2\times10^4$K, $\zeta=20$, and is in agreement with both constraints.  As $\Tvir$ is increased, reionization occurs more rapidly and at later times.  To fit the upper limits on $\avenf$, this increase in $\Tvir$ must be accompanied by an increase in $\zeta$.  This is illustrated by the green dot-dashed curve, which corresponds to the corner of our parameter space, $\Tvir=3\times10^5$K, $\zeta=50$.  Increasing $\Tvir$ also steepens the redshift evolution of the emissivity.  Therefore, even though one can increase the ionizing efficiency to compensate for the paucity of sources, the evolution of the emissivity becomes too steep to match the data at $3<z<6$, in the absence of powerful and rapid feedback mechanisms or an equally rapid evolution in $\zeta$ post reionization.  This is illustrated by the dashed red curve, corresponding to $\Tvir=10^6$K, $\zeta=300$.
\label{fig:emissivity}
}
\vspace{-1\baselineskip}
\end{figure}

However, the tighter SPT bound, $\Ptot<2.8~\muKK$ at 95\% C.L., {\it is incompatible with all of our reionization models.} Smaller patchy reionization signals than our predicted range, i.e. $\PkSZ < 1\muKK$ -- as would accommodate this SPT constraint -- require very rapid and late reionization histories (i.e., spanning $z\sim 6-7$).  In these, ionizing sources would reside in extremely rare, massive halos ($\Tvir\gsim10^6$ K) such that their abundance grows rapidly with time.
  However, such models are difficult to accommodate with current constraints:  (i) transmission in the Lyman forests of high-$z$ QSOs provides a low-$z$ limit to reionization so that reionization cannot occur too late (e.g., \citealt{MMF11}); (ii) the redshift evolution of the transmission in the Lyman forests favors a slow evolution in the ionizing emissivity and, thus, a more extended reionization process \citep{Miralda-Escude03, BH07}; (iii) the luminosity function of high-redshift galaxies (e.g., \citealt{Bouwens08}) supports a sizable contribution to reionization by galaxies fainter and more abundant than those observed\footnote{Note that the observed galaxies roughly correspond to the largest values of $\Tvir \sim$ few $\times10^5$ K used in this study (e.g., \citealt{Labbe10, SFD11, FDO11}).}. To illustrate points (i) and (ii) in greater detail, we plot the redshift evolution of the ionizing photon emissivity and $\avenf$ in Fig. \ref{fig:emissivity}.  Increasing $\Tvir$ steepens the redshift evolution of the emissivity.  Therefore, even though one can increase the ionizing efficiency to compensate for the paucity of sources (appealing to larger than expected values of $f_\ast f_{\rm esc}$ or more exotic, top heavy stellar populations), the evolution of the emissivity becomes too steep to match the data at $3<z<6$, in the absence of a rapid counter-evolution of $\zeta$ or rapid feedback mechanisms.  This is illustrated by the dashed red curve in Fig. \ref{fig:emissivity}, corresponding to $\Tvir=10^6$K, $\zeta=300$.

   If the more aggressive SPT bound is confirmed, more exotic reionization scenarios that reduce the level of patchiness may become favorable.  The most plausible candidates would be very high-energy X-rays with long mean free paths, which would significantly ``pre-ionize'' the IGM (we show above that the early to mid stages of reionization are the most relevant for the $l=3000$ signal).   A homogeneous ionization floor of amplitude $x_i$  suppresses the kSZ power by the factor $\sim (1-x_i)^2$ (see \citealt{VL11} for a simple model of X-ray reionization and the kSZ).  For example, if X-rays homogeneously pre-ionize the IGM to $x_i=0.3$, the patchy signal from our models would decrease to $\PkSZ \lsim$ 0.7--1.7 $\muKK$, with some models accommodating the aggressive SPT bound.  The patchy epoch of reionization would also occur quicker given a homogeneous head-start, further decreasing the kSZ signal.  However, a homogeneous pre-ionization is also an extreme simplification, as X-rays would still imprint some inhomogeneities on angular scales of $l=3000$ ($\sim 20$ Mpc at high redshifts), given that the majority of X-ray photons have mean free paths much less than the Hubble length even for hard QSO spectra.

In any case, since late-forming massive halos host both hot gas and star forming galaxies, it is likely that the CIB and the tSZ are correlated (e.g., \citealt{Sehgal10}).  This would invalidate the aggressive $\Ptot< 2.8~\muKK$ constraint proposed by \citet{Reichardt11}, implying that the more conservative SPT bound of $\Ptot < 6~ \muKK$ is more plausible.  This limit is consistent with all of our reionization scenarios, and is on the cusp of the probing early/extended reionization models.  A $\approx1~\muKK$ constraint on the signal is projected for the coming year \citep{Reichardt11}, which should result in the first detection of the kSZ.

\section{Conclusions}
\label{sec:conc}

Observational efforts such as ACT and SPT are beginning to set interesting constraints on the kSZ power spectrum at $l\approx3000$.  This is highlighted by the recent SPT constraint of $\Ptot< 2.8~\muKK$ (\citealt{Reichardt11}; $<6~\muKK$ when allowing for correlations), as well as the ACT measurement of $\Ptot =  6.8\pm 2.9~\muKK$ \citep{Dunkley11}.  A $\approx1~\muKK$ constraint on the signal is projected for the coming year \citep{Reichardt11}, which should result in the first detection of the kSZ.  Second generation instruments, such as ACTPol and STPPol, will further improve the kSZ measurements in the coming years.  

This study calculated the patchy kSZ signal for a suite of $\approx100$ inhomogeneous reionization scenarios, which were generated using the publicly-available code \cmfast.  
In these models, the contribution of inhomogeneous reionization to the CMB angular power spectrum at $l=3000$ spans only a factor of four, $\PkSZ \approx$ 1--4 $\muKK$, despite the large volume of parameter space covered by the reionization models. 
The allowed range narrows further to $\PkSZ\approx$ 1.5--3.5 $\muKK$ when including current constraints on the reionization history from {\it WMAP} and high-redshift quasars.  We find that for standard models, the bulk of the signal is imprinted in the early to middle stages of reionization, when the (clustered) ionized structures have angular scales comparable to $l\approx3000$.

 The range $\PkSZ \approx$ 1.5--3.5 $\muKK$, combined with estimates for the post reionization kSZ of $\POV \approx$ 2--3 $~\muKK$, results in a total kSZ signal of $\Ptot = $3.5--6.5 $\muKK$.  Thus, the recent conservative SPT bound of $\Ptot <6~\muKK$, which assumes a free, scale-independent CIB-tSZ correlation, is on the boarder of constraining reionization, even assuming conservatively low values for $\POV$.  The strongest and easiest to detect patchy kSZ signals correspond to early and extended reionization scenarios.  In such models, the sources of ionizing photons are abundant at early times and/or there are many recombinations (absorptions in sinks) so that reionization was maximally extended.  Furthermore, with the assumption of a slope (but not amplitude) in the CIB-tSZ correlation, \citet{Zahn12} placed an intermediate upper limit of $\Ptot \lsim 4 \muKK$.  If such a slope is credible, this constraint is sufficient to rule out our reionization models driven by molecularly-cooled halos (with $\Tvir\lsim10^4$ K), and place limits of $\zre\lsim10$, $\delz\lsim2$ (see Fig. \ref{fig:zre_delz}).

On the other hand, the tightest SPT bound $<2.8~\muKK$ at 95\% C.L., which assumes no CIB-tSZ correlation, is incompatible with {\it all} of our reionization scenarios.  Although later and more rapid reionization models might be in agreement with this constraint, it is difficult to push our models to such extremes since these would conflict with other observations of the high-redshift Lyman alpha forest and galaxy luminosity function (LF). This implies that either: (i) the early to middle stages of reionization occurred in a much more homogeneous manner than suggested by the stellar-driven scenarios we explore, such as would be the case if, e.g., very high energy X-rays or exotic particles contributed significantly; and/or (ii) that there is a significant correlation between the CIB and the tSZ, which invalidates these bounds.  Upcoming combined analyses of microwave and Herschel far infrared data will likely test the later possibility.  

  We also find that shape of the patchy kSZ power spectrum encodes astrophysical information.  An early and extended reionization in which bubble growth is retarded by recombinations results in a kSZ power spectrum which peaks on smaller scales. The sign of the slope of the patchy kSZ power spectrum at $l\approx3000$ is useful for discriminating between extended/early and rapid/late reionization processes, which have the same $\PkSZ$.  In other words, the values of $\tau_e$ and $\PkSZ$ do not uniquely determine the redshift and duration of reionization (c.f., \citealt{Zahn12}).  However, the sign of the kSZ slope at $l\sim3000$, combined with $\tau_e$ and $\PkSZ$, is sufficient to determine the redshift and duration of reionization. Furthermore, we find that positive values of the slope come exclusively from models with low values of $\mfp \lsim 10$ Mpc.  Therefore detecting the sign of the slope can constrain the abundance of absorption systems in the early Universe, at redshifts far beyond those of present estimates based on QSO spectra.

We anticipate that these measurements will be complemented by analyses of the Planck EE signal, which is also sensitive to the duration of reionization, and potentially by analysis of the contribution of patchy reionization to the CMB polarization power spectrum.  Reionization probes not based on the CMB will also yield complementary information, using 21cm tomography, QSO spectra and wide-field LAE studies. These tools will soon reveal the details of the dawn of cosmic structure.



\vskip+0.5in

We especially thank Hy Trac for useful discussions, and Andrea Ferrara for comments on an early draft of this paper.
We thank Dave Spiegel for his considerable Matlab assistance without which most of the figures presented here would have been much less visually appealing. AM would like to thank the hospitality of Princeton University where most of this work was completed. MM is supported by NASA through an Einstein Fellowship.

\bibliographystyle{mn2e}
\bibliography{ms}

\appendix
\section{Convergence Tests}
\label{sec:appendix}

\begin{figure}
\includegraphics[width=0.45\textwidth]{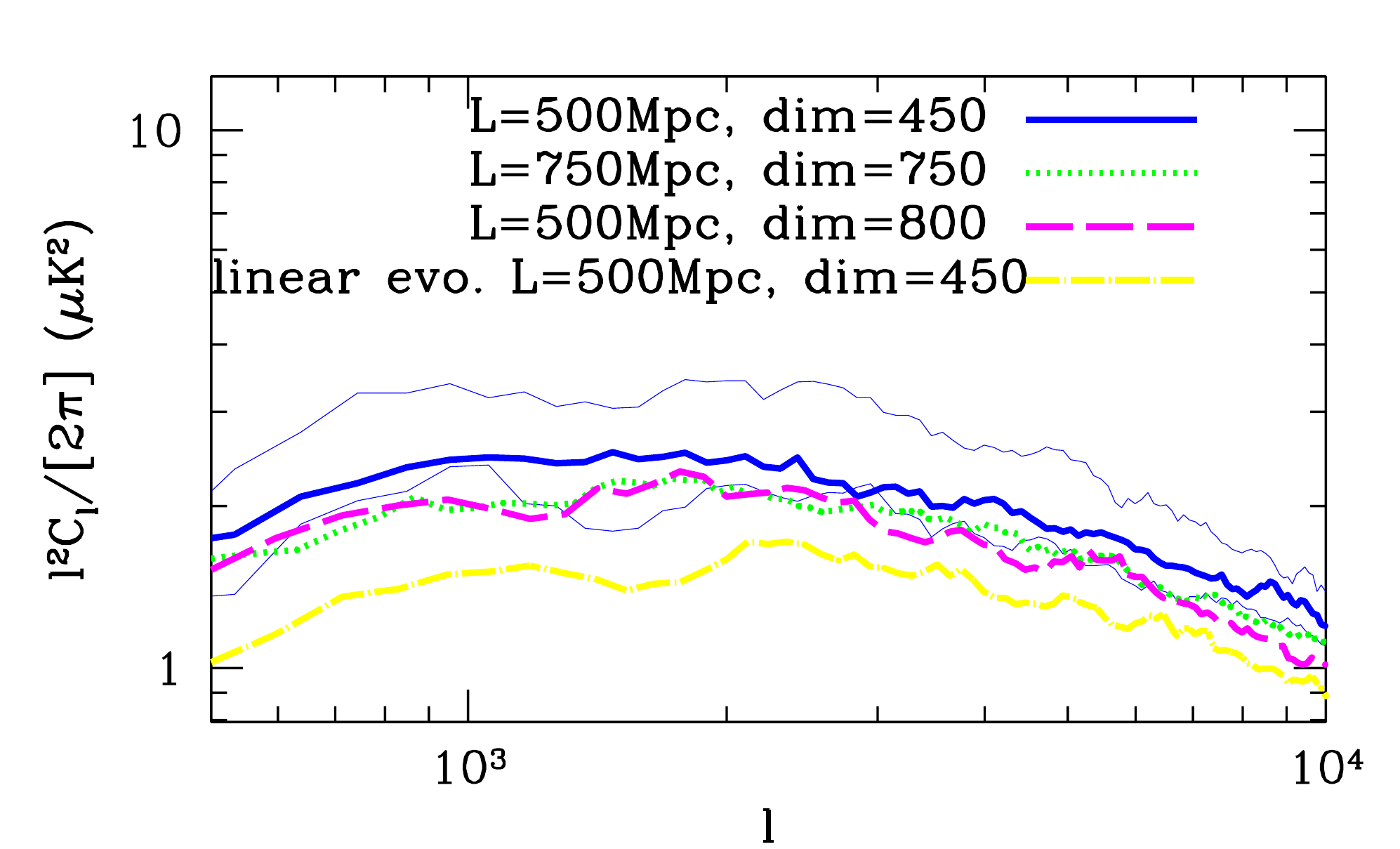}
\caption{
The patchy kSZ power spectra obtained from different simulations.  The (thick) solid blue curve corresponds to the same box size and resolution as used in this work: $L=500$ Mpc on a side, sampled onto a grid with dimensions given by dim=450.  The dotted green curve corresponds to a larger box, with comparable resolution: $L=750$ Mpc, dim=750.  The dashed magenta curve has the same box size as our fiducial choice, but with higher resolution, dim=800.
  The thin solid blue curves show power spectra computed along two basis vectors without rotating the simulation box, for our fiducial choices.  
  Finally, the dot-dashed yellow curve corresponds to the signal computed without PT, just assuming linear evolution of the cosmological fields, such as has been done in previous semi-analytic studies.  
  All curves assume $\reionparams$ = \{32, 10$^4$ K, 30 Mpc\}.
\label{fig:convergence}
}
\vspace{-1\baselineskip}
\end{figure}

Here we briefly quantify the convergence of our numerical results.  In Fig. \ref{fig:convergence}, we plot the patchy kSZ power spectra corresponding to the same astrophysical parameter choices, $\reionparams$ = \{32, 10$^4$ K, 30 Mpc\}, but with different simulation box sizes, $L$, and grid dimensions, dim. The (thick) solid blue curve corresponds to the same box size and resolution used in this work.  The dotted green curve corresponds to a larger box, with comparable resolution, while the dashed magenta curve has the same box size as our fiducial choice, but with higher resolution.  Comparing the blue curve to the magenta and green ones, suggests that our fiducial choices result in a slight overestimate of kSZ power spectrum for this random seed, but have converged to within 10\% over the relevant multipoles.

  The thin solid blue curves in Fig. \ref{fig:convergence} show power spectra computed along two basis vectors without rotating the simulation box.  From their disparity, one can see that the variance of the velocity fields on scales larger than our $L=500$ Mpc scales is 10\% (Fig 7, \citealt{Iliev07kSZ}), requiring box rotations to avoid coherently stacking the signal.  Finally, the dot-dashed yellow curve corresponds to the signal computed without PT, just assuming linear evolution of the cosmological fields, such as has been done in previous semi-analytic studies \citep{McQuinn05}.  Ignoring non-linearity results in a $\sim$ 30\% underestimate of the $l\sim$ 2000--10,000 kSZ power in this model.

\begin{figure}
\includegraphics[width=0.45\textwidth]{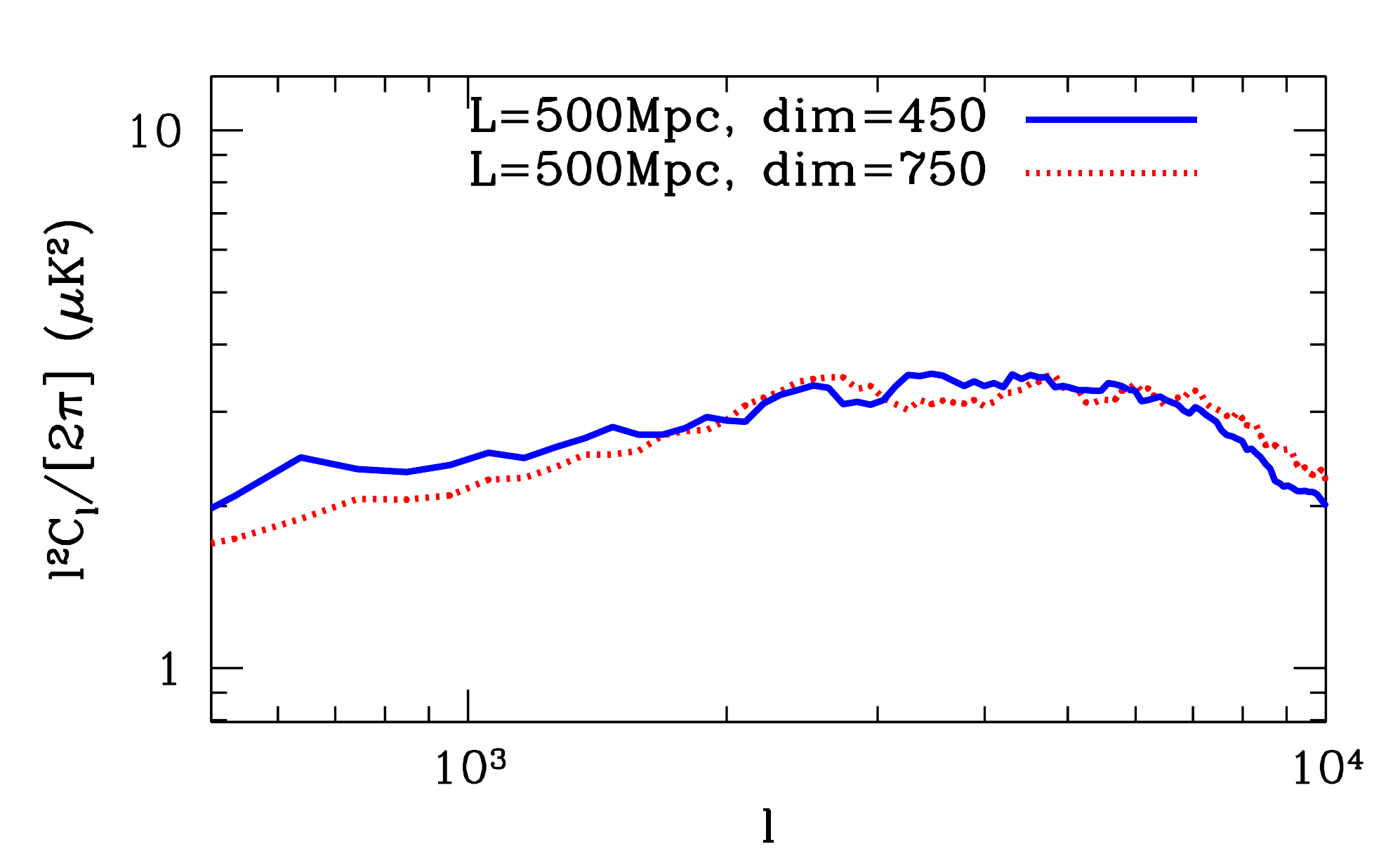}
\caption{
\label{fig:mfp3_convergence}
The kSZ signal corresponding to $\reionparams$ = \{32, $6.3\times10^3$ K, 3 Mpc\} from our fiducial simulation ({\it solid blue curve} and a higher resolution, dim=750 simulation ({\it dotted red curve}).
}
\vspace{-1\baselineskip}
\end{figure}

One might also worry that our fiducial resolution, with a cell size of 1.1 Mpc, is insufficient to accurately capture the kSZ signal from our low-$\mfp$ models. To test whether this resolution limit has a quantitative impact on our results, we ran a higher resolution, dim=750, run with the same astrophysical parameters as the model shown in the bottom panel of Fig. \ref{fig:ionization_maps}: $\reionparams$ = \{32, $6.3\times10^3$ K, 3 Mpc\}.  We compare their kSZ power spectra in Fig. \ref{fig:mfp3_convergence}, and find that the signal has converged.

\end{document}